\documentclass[10pt,journal,compsoc]{IEEEtran}
\usepackage{fancyhdr}

\ifCLASSOPTIONcompsoc
  \usepackage[nocompress]{cite}
\else
  \usepackage{cite}
\fi
\ifCLASSINFOpdf
  \usepackage[pdftex]{graphicx}
\else
\fi
  \usepackage[caption=false,font=footnotesize,labelfont=sf,textfont=sf]{subfig}
\usepackage{hyperref}
\hypersetup{
    colorlinks=true,
    linkcolor=black,
    citecolor=black,
    urlcolor=black
}
\usepackage{cleveref}

\usepackage{todonotes} 

\clearpage
\begin{document}

%
\title{Words of Estimative Correlation: Studying Verbalizations of Scatterplots}

%
%
%
%

\author{Rafael Henkin and Cagatay Turkay~\IEEEmembership {Member,~IEEE}%
\IEEEcompsocitemizethanks{\IEEEcompsocthanksitem 
 Rafael Henkin performed this work while with the giCentre at City, University of London and is now with the Centre for Translational Bioinformatics, Queen Mary, University of London. E-mail: r.henkin@qmul.ac.uk %
 \IEEEcompsocthanksitem Cagatay Turkay is with the Centre for Interdisciplinary Methodologies, University of Warwick, UK. E-mail: cagatay.turkay@warwick.ac.uk}%
\thanks{Manuscript received April 19, 2019; revised August 26, 2019.
}%
}

%
%

\markboth{Journal of \LaTeX\ Class Files,~Vol.~14, No.~8, August~2015}%
{Shell \MakeLowercase{\textit{et al.}}: Bare Demo of IEEEtran.cls for Computer Society Journals}
%



\IEEEtitleabstractindextext{%
\begin{abstract}
Natural language and visualization are being increasingly deployed together for supporting data analysis in different ways, from multimodal interaction to enriched data summaries and insights. 
Yet, researchers still lack systematic knowledge on how viewers verbalize their interpretations of visualizations, and how they interpret verbalizations of visualizations in such contexts.
We describe two studies aimed at identifying characteristics of data and charts that are relevant in such tasks. The first study asks participants to verbalize what they see in scatterplots that depict various levels of correlations. The second study then asks participants to choose visualizations that match a given verbal description of correlation. We extract key concepts from responses, organize them in a taxonomy and analyze the categorized responses. We observe that participants use a wide range of vocabulary across all scatterplots, but particular concepts are preferred for higher levels of correlation. A comparison between the studies reveals the ambiguity of some of the concepts. We discuss how the results could inform the design of multimodal representations aligned with the data and analytical tasks, and present a research roadmap to deepen the understanding about visualizations and natural language.


\end{abstract}

}

\maketitle
\thispagestyle{fancy}

\IEEEdisplaynontitleabstractindextext

%
\IEEEpeerreviewmaketitle

\IEEEraisesectionheading{\section{Introduction}\label{sec:intro}}









Recent advances in machine learning and artificial intelligence have paved the way for novel multimodal interactive methods that mix natural language and visualization to facilitate discovery and improve workflows within data analysis~\cite{Hoque2017, yu2019flowsense, fast2018iris}. Natural language statements can be used to complement the information that is displayed visually, for example, by reducing clutter with statistical summaries~\cite{sevastjanova2018going}, by providing complementary information in textual form that are not perceived easily with graphics~\cite{Demiralp2017}, or as in the case of automated reports, by contextualizing visual summaries and charts within a narrative~\cite{Mumtaz2019}. 
There is, however, a lack of empirical research on understanding how viewers verbalize their interpretations of visualizations and how they interpret verbalizations of visualizations in data analysis contexts. This paper is motivated by this gap and aims to develop a broader understanding on the data and chart characteristics that are of significance when people view, interpret, and describe data visualizations. 


Our goal here is to establish empirical evidence that can inform decisions in designing and developing multimodal data analysis approaches. The role of such empirical knowledge bases are reported in both of the areas that our work relates to -- visualisation and natural language processing. In visualisation, perceptual studies~\cite{Harrison2014,Rensink2011} shed light into how particular patterns are perceived and which visual representations work better for particular tasks. In natural language generation, system designers seek to acquire such empirical data in the form of ''\textit{relevant knowledge about the domain, the users and the language used in texts}''~\cite{Reiter2003} within the \textit{knowledge acquisition} stage. This process ensures that natural language statements produced by systems are aligned with the domain of supported analytical tasks. 

This knowledge need can be addressed for specific application contexts as seen in existing case studies~\cite{Setlur2016,Srinivasan2017a}. However, we posit that an application-agnostic approach, influenced by methods from natural language processing (NLP) and cognitive science literature, enables forming a greater understanding on the combination of visualization and natural language across contexts. In pursuing this vision, we follow the examples of experiments studying language in cognitive psychology where, for instance, the inference of comparative words such as ``smaller'', ``larger'' are studied~\cite{skylark2018says}, or where the perception of terms relating to probabilities are investigated, as in the seminal ``\textit{Words of Estimative Probability}'' work by Sherman Kent and colleagues~\cite{kentsherman1964} -- which also inspires the title of this paper.



To that end, this paper reports a descriptive and relational investigation~\cite{Rosenthal1984} through the design and analysis of two studies focusing on the exploration of correlation relations as visualized through scatterplots. We gather empirical data on how people describe scatterplots of varying correlation and how they map scatterplots to given textual descriptions of correlations. We choose to focus on correlation since it is one of the most fundamental analysis tasks in data science~\cite{james2013introduction}, and on scatterplots as they are the most effective and commonly used method to explore correlations~\cite{Harrison2014}.
The design and analysis of the studies are motivated by three overarching goals: \textbf{G1} - gaining an overview of the language constructs and the vocabulary used to describe the visualizations, \textbf{G2} - identifying the characteristics of the data and the visualizations that participants refer to, and \textbf{G3} - understanding the relationship between these characteristics and the language observed in the studies.
 



\textit{In the first study}, we examine how variations in the underlying correlation relations affect the descriptions of scatterplots. More specifically, we ask participants to \textit{verbalize} what they see in different scatterplots. Through an analysis methodology that comprises a combination of NLP techniques and qualitative analysis of responses, we extract key characteristics of data and visualization. We use techniques that break down the responses into sequences of words, i.e., \textit{syntactic analysis} step of natural language processing~\cite{manning1999foundations}, and identify the relationships between words from a linguistic perspective. We then follow a qualitative approach drawing on the thematic analysis methodology~\cite{braun2006using} and organize the words under a taxonomy comprising five \textit{concepts} -- the different characteristics that participants refer to -- and five \textit{traits} -- the qualities or quantities of these characteristics.

We categorize and analyze the responses, revealing a wide range of vocabulary used by participants to describe the relations, with responses often involving multiple concepts and traits with varying relations to correlation values. We observe, for example, that descriptions referring to the strength of the relationship are used uniformly across the whole range of correlations, while descriptions referring to the direction and orientation of the patterns in the plots are more common for a narrow range of correlation values. 


\textit{In the second study}, we reverse the experiment by showing constructed verbalizations to participants and asking them to choose the scatterplots that they think represent those verbalizations. With this study, we are able to identify the extent to which the verbal descriptions resonate with viewers for varying levels of correlation. To gain further insights on the relationship between the responses and the categories of the taxonomy, we also compare the results of the two studies, starting to reveal potential differences in the processes of \textit{producing} and \textit{interpreting} verbal descriptions of visualizations. The comparative analysis between the first and the second experiments' results reveals differences in the interpretation of some word combinations, with descriptions of orientation, for example, being less ambiguous than descriptions of the absolute existence of a relationship.

We then reflect on the observations made from the two studies and discuss the potential implications of the results in designing multimodal systems and visualisations of correlations. We also discuss the potential of our study and analysis methodology for furthering research in this area, as well as reflecting on the limitations. This paper's contributions can be described as follows:
\begin{itemize}
    \item empirical data on how people interpret and verbalize correlation in scatterplots, a taxonomy that reveals the high-level characteristics within the responses, and distributions of the responses over the categories for varying levels of correlation. The combination of these provide insights into how correlation relations are verbalized and constitutes a knowledge base to construct reliable descriptions of visual representations (sections~\ref{sec:findings-s1}, ~\ref{sec:s2resultsanalysis});
    \item a methodological blueprint comprising the study protocol and a semi-automated analysis approach that can be utilized for further empirical studies exploring the relation between natural language and visualization;
    \item a research roadmap along the goal of understanding how natural language and visualization work together in data analysis contexts (section~\ref{sec:roadmap}).
\end{itemize}

\section{Related work}
\label{sec:relatedwork}

\subsection{Visualization and natural language}


The use of natural language generation (NLG) in complementing visual information has been investigated in several application areas, as well as in domain-agnostic settings. As authors often describe the development of multimodal systems as a whole, the language realization step is rarely discussed, with issues about the use of certain words often appearing during evaluation. Srinivasan et al.~\cite{Srinivasan2019} combine the generation of data-driven statements with the exploration of alternative visualizations to facilitate visual analysis. Study participants report the need to understand what the system meant when using words such as \textit{moderate} and \textit{strong}. Mumtaz et al.~\cite{Mumtaz2019} introduce a multimodal tool to help with code quality analysis and acknowledge this by including tooltips to clarify the boundaries when values change from \textit{low} to \textit{medium}. A slightly different approach, less focused on natural language, was used by Demiralp et al.~\cite{Demiralp2017} when providing an \textit{insight}-based visualization exploration system.  Jain and Keller~\cite{Jain2015} also report changing their textual health alerts based on feedback for words such as \textit{significant}. There are also instances of authors using synonym lists to keep statements interesting~\cite{Latif2019}, but in none of these cases there is a thorough consideration of the alignment of data, visualizations and natural language from a linguistic perspective. The only similar works are in narrowing down color names across multiple languages~\cite{Heer2012,Kim2019}.


Natural language is also used as an input in multimodal systems, as a means to provide alternative methods for hands-free interaction~\cite{Srinivasan2017a,Srinivasan2017} or as a complementary medium to traditional interaction modes~\cite{Gao2015,Hearst2019a,Hearst2019b,Hoque2017,Setlur2016,Sun2010}, allowing users to communicate with their own words rather than learning a potentially complex set of interactions. More complex products such as Tableau\footnote{http://www.tableau.com} include both the ability of users to use text as input and the use of data-driven textual statements as outputs; tools like Quill\footnote{https://narrativescience.com/quill} also facilitate the creation of data-driven stories to complement data analysis software. The studies we present here also have a complementary role for such these cases, helping to reveal the different strategies that people use to communicate about visualizations and that can then be adapted accordingly in tools. 


\subsection{Perceptual studies}

Our work is also related to the investigation of perception of visualizations. Researchers have long sought to investigate how humans perceive correlation in scatterplots, with various studies~\cite{Bobko1979,Pollack1960} exploring the ability to estimate varying levels of correlation. More recently, the information visualization community has approached this problem from a design perspective, focusing on the perception of correlation as a way to model the effectiveness of visualization design. Results from these studies showed how variation of correlation for normally distributed variables can be modeled with a few parameters~\cite{Rensink2011}, and also systematically demonstrated how different types of visualization are more or less suitable for these tasks~\cite{Harrison2014,Kay2016}. Beecham et al.~\cite{Beecham2017} extended such studies to inferences from visual representations of geospatial data derived from systematic variations of underlying spatial structure. Sher et al.~\cite{Sher2017} reported results that brought new questions to the field, demonstrating the difficulty of estimating correlation accurately when the underlying data deviated from normal bivariate distributions. Correll and Heer~\cite{Correll2017} explored similar conditions when investigating how reliable are people's estimates of trends and missing values. The idea of systematically presenting varied levels of correlation, common in these other works, served as an inspiration for our studies. Some of the variations in data distributions are also part of our research agenda. Our paper, nonetheless, investigates the highly subjective use of natural language rather than the accuracy of estimations.  

Other tasks were also tested regarding the depiction of correlation with scatterplots. Pandey et al.~\cite{Pandey2016} investigated the perception of scatterplots based on judgment of similarity. Rather than modeling perception based on correlation, their aim was to identify the perceptual features of scatterplots used by participants to group them. The output of the study is a condensed list of concepts that describe scatterplots. Their results consist mostly of references to characteristics of the distribution of points, which partially overlaps with our findings. Our taxonomy, however, includes a wider range of characteristics.
\section{Study S1: Visualization to Verbalization}
\label{sec:e1}

The first study consisted of asking participants to provide textual descriptions -- \textit{verbalizations} -- of scatterplots. The result is a collection of responses, or \textit{utterances}, distributed by the varying levels of correlation that are displayed through scatterplots. In this section we describe the preparation of materials and the study procedure, details about the analysis of the responses and present and discuss the derived taxonomy. For stimuli, survey, collected responses and analysis code, see the online repository, located at \url{https://github.com/nlvis/wec}.

\begin{figure}[!t]
\centering
\includegraphics[width=.80\columnwidth]{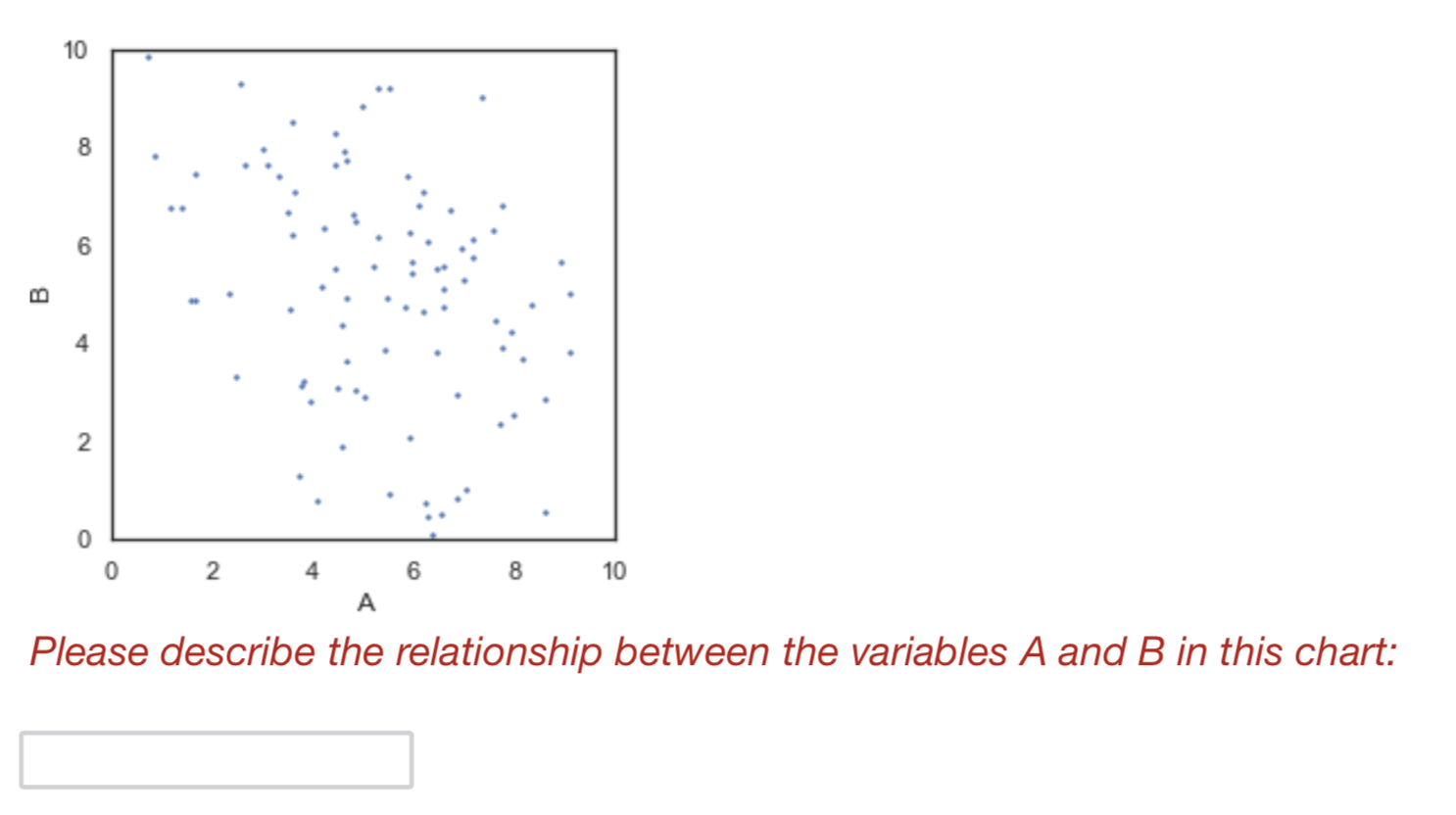}
\caption{Screenshot from user study 1 where we ask participants to describe the relation in a scatterplot.}
\label{fig:os1}
\end{figure}

\subsection{Preparation}



\begin{figure}
\centering
\includegraphics[width=.95\columnwidth]{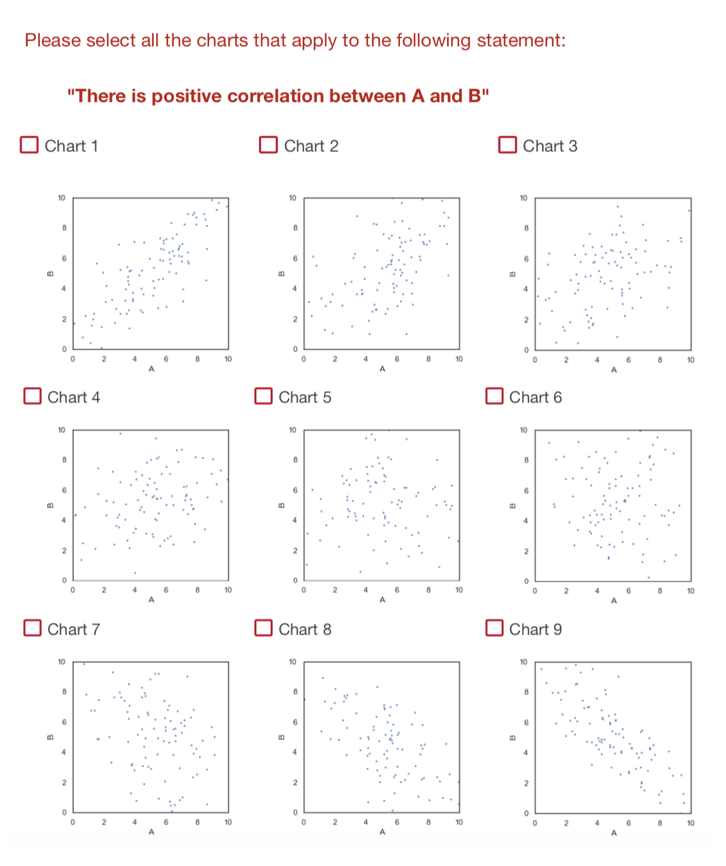}
\caption{Screenshot from user study 2 where we ask participants to indicate the scatterplots that match a verbal description of a given relationship. The nine scatterplots here are the same as in study 1.}
\label{fig:os2}
\end{figure}

\noindent \textbf{Materials and participants:} We generated 9 scatterplots (see \figurename~\ref{fig:os2}) inspired by the scatterplots used in previous research~\cite{Rensink2011,Harrison2014,Kay2016}. They were generated by drawing samples from a bivariate normal distribution, with correlations ranging from -0.8 to 0.8, in increments of 0.2. Scatterplot figures are 288 pixels in both dimensions with 100 equal-sized points each, axes named after abstract variables \textit{A} and \textit{B}, and with ticks drawn for even numbers from 0 to 10. We used abstract variable names (\textit{A} and \textit{B}) instead of real concepts (such as birth rate, GDP, etc.) for two reasons: first to avoid biasing the responses due to participants relating more (or less) to the data context, and second, to gather as much context-agnostic language as possible in order to simplify the text analysis and the interpretation phase. For the study, we recruited 160 participants through the Prolific platform\footnote{https://www.prolific.co/} and our only pre-screening filter was self-declaration of native language as English. 

\medskip


\noindent \textbf{Procedure:} Each participant was asked to write an answer to the instruction \textit{``Please describe the relationship between the variables A and B in this chart''} for each scatterplot, in a randomized order. Before proceeding with the task, participants were presented with a description of how scatterplots are used in data analysis -- we note that this description included the word \textit{relationship}, but did not include the term \textit{correlation}. We carefully avoided biasing participants in this training by not providing any interpretation or description of the relationship between the variables, and instructed participants to give complete answers and avoid answering with \textit{same as before} or \textit{same}. 

\subsection{Results and analysis}
We collected 1428 responses from 160 participants. One ``n/a'' answer was removed, and an early run had 11 participants seeing 8 scatterplots instead of 9 due to an error in the survey. Since we aggregate all the responses and are not exploring individual differences in this study, we opted for keeping these responses in. During the analysis, we combine natural language processing (NLP) methods with manual coding (See Section~\ref{sec:discussion} for a reflection on this decision). We decided not to use classification or clustering techniques such as topic modeling since our aim was to map concepts to answers rather than classify answers by concepts. In this section, we detail the use of NLP techniques to process the responses, followed by the derivation of the concepts and traits. As mentioned, all the steps described in the paper are in the Python notebooks available in the online repository.\\
\subsubsection{NLP steps}

\noindent \textbf{Pre-processing:} as part of the data cleaning process, we corrected spelling errors such as ``realtionship'' with the aid of a spell checker. We did not change words such as ``correlationship'', as it is not clear how to correct such cases without asking the participants. We also replaced occurrences of references to variables, such as \textit{variable A}, with words \textit{VarA} and \textit{VarB} in order to make such references easy to identify in subsequent analysis.\\

\noindent \textbf{Syntactic analysis:} to extract the main ideas from answers, we used NLP techniques that break the responses down and identify the role of individual words in sentences. This process is called \textit{parsing} or \textit{syntactic analysis}, and involves several algorithms and methods~\cite{manning1999foundations}. The first step, \textit{tokenization}, breaks text down into multiple sentences, and sentences into individual words -- the tokens. Here, we used \textit{regular expressions} that preserve hyphenated words. The second step is \textit{part-of-speech} tagging and involves categorizing the words, now separated as \textit{tokens}, based on the grammar of a specific language and how they appear in the sentence. This means, for example, identifying if a word such as ``increase'' is being used as a verb or as a noun in a sentence. In our work, we used the Stanford CoreNLP library~\cite{Toutanova2007}, which employs a state-of-the-art model for the English language. Finally, the \textit{lemmatization} step groups inflected forms of words together, i.e., plurals, past and future tenses, etc. With this, words such as ``am'', ``is'' and ``are'' are mapped to the infinitive verb ``be''. This was done using the WordNet lemmatizer~\cite{Bird2009}, a model that works for words identified as nouns, adjectives and verbs. These steps ensure that it is possible to analyze and compare responses independent of participants' sentence construction preferences, focusing instead on the words and their meanings. The last step of the syntactic analysis was the removal of ``stop words'' from each response. These are words that are not particularly relevant to the analysis and include prepositions and articles (e.g., ``and'', ``the'', etc.).\\


\begin{figure}
\centering
\includegraphics[width=.7\columnwidth]{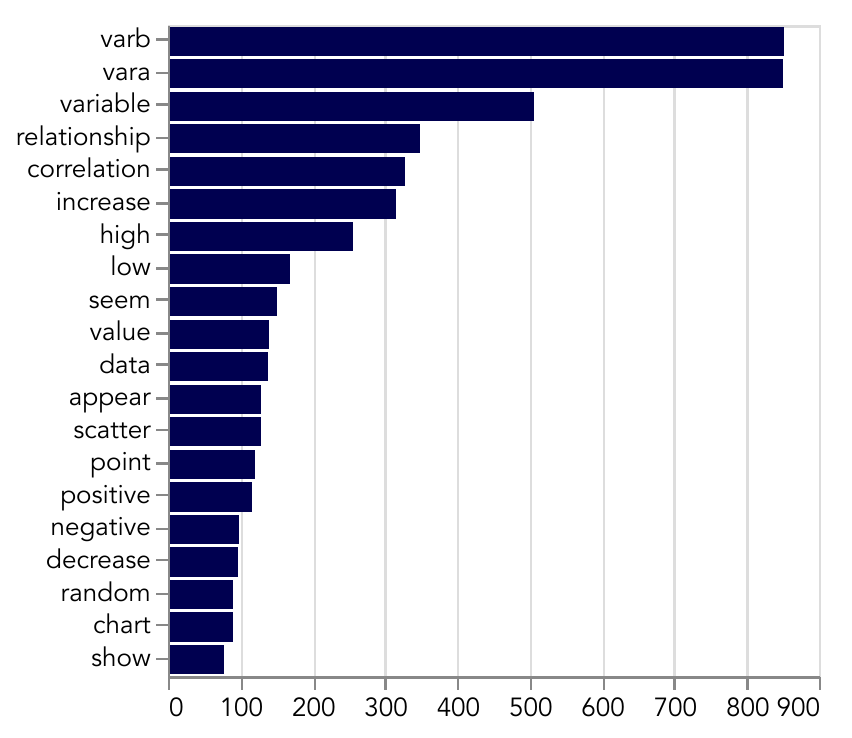}
\caption{Top 20 words extracted from all answers; nouns ``A'' and ``B'' were replaced by \textit{vara} and \textit{varb} to facilitate analysis.}
\label{fig:topwords}
\end{figure}

\begin{table*}[!htbp]
\centering
\caption{Definition, words and examples of answers of the concepts identified through the analysis of results.}
\includegraphics[width=.85\textwidth]{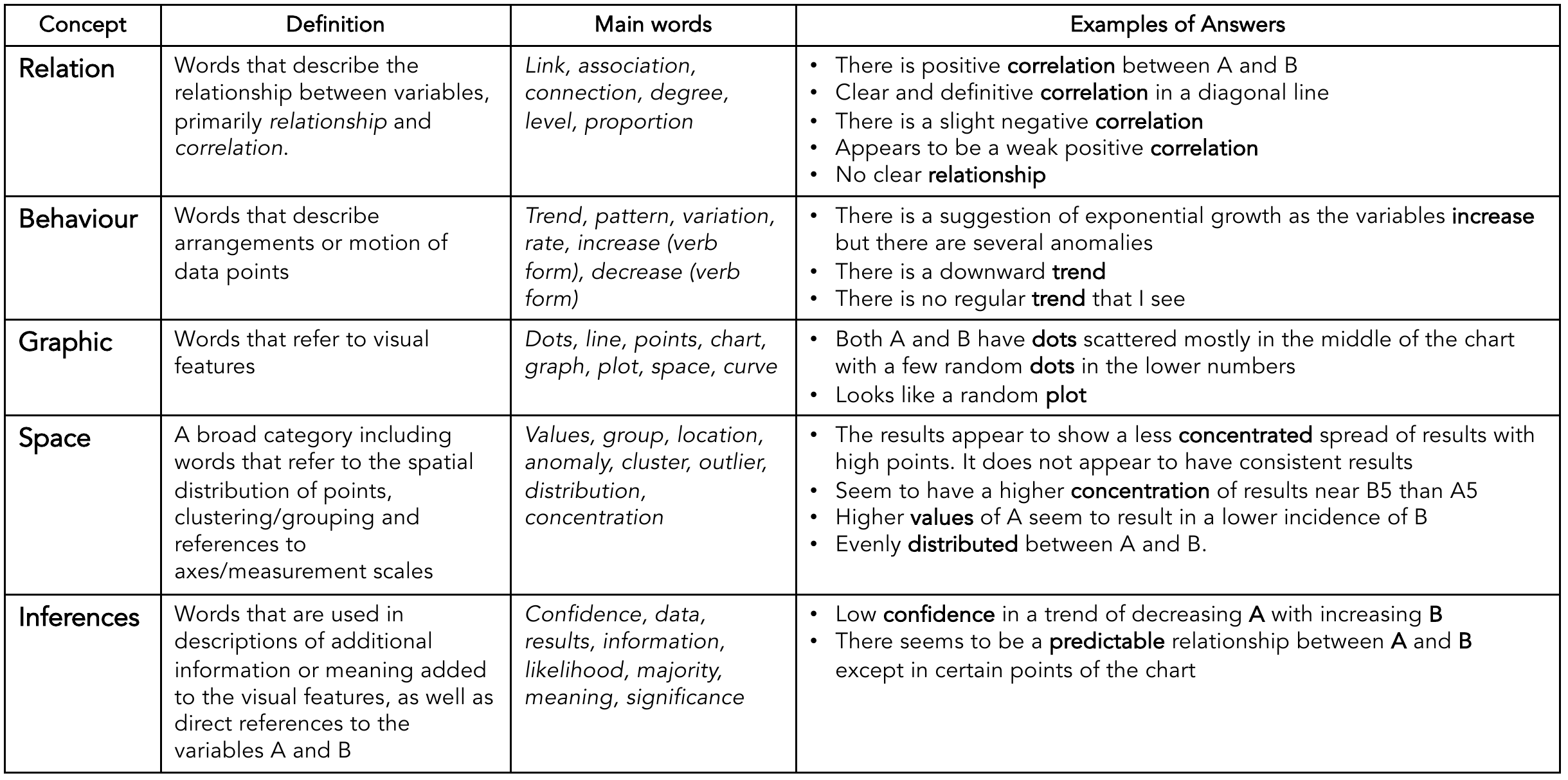}
\label{fig:table-concepts}
\end{table*}

\begin{table*}[!htbp]
\centering
\caption{Definition, words and examples of answers for the traits identified through the analysis of results.}
\includegraphics[width=.85\textwidth]{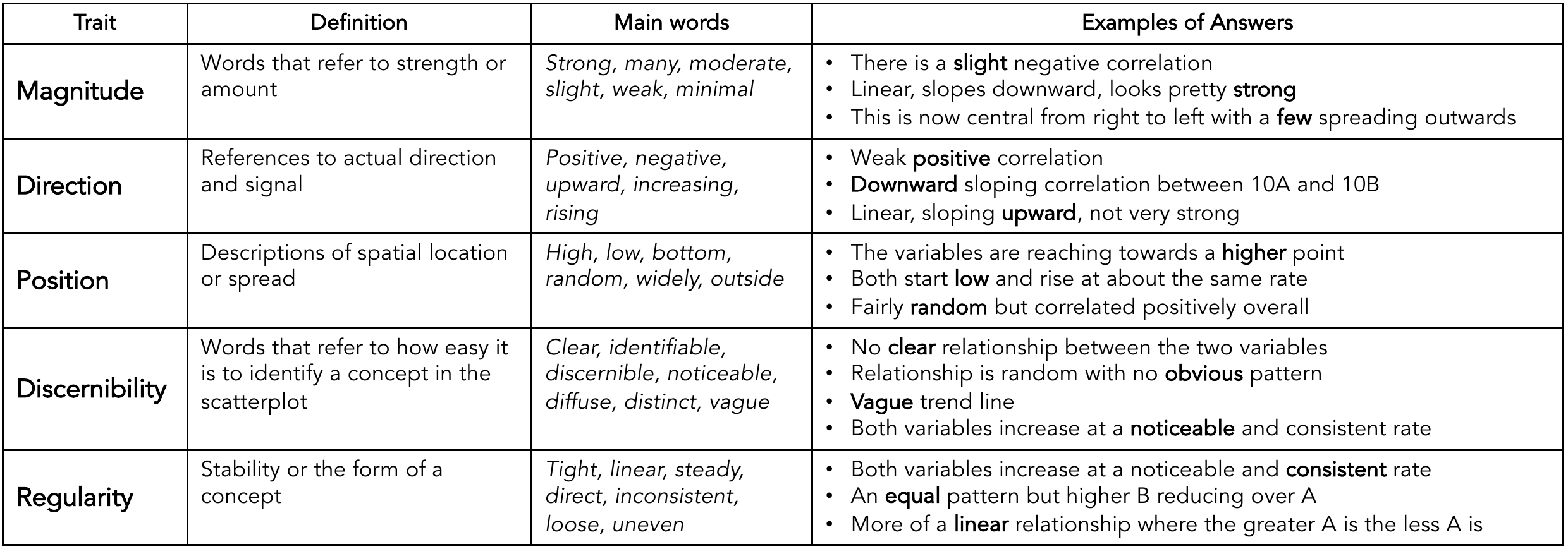}
\label{fig:table-traits}
\end{table*}

\begin{table*}[!htbp]
\centering
\caption{Most common combinations of concepts, traits and negative statements with examples. With the exception of one combination that includes 5   concepts or traits, most other common combinations are for 2 or 3 concepts or traits. For the bottom four combinations in the table, the lower number of answers reinforces the notion that participants used a wide variety of unique combinations of concepts and traits.}
\includegraphics[width=.8\textwidth]{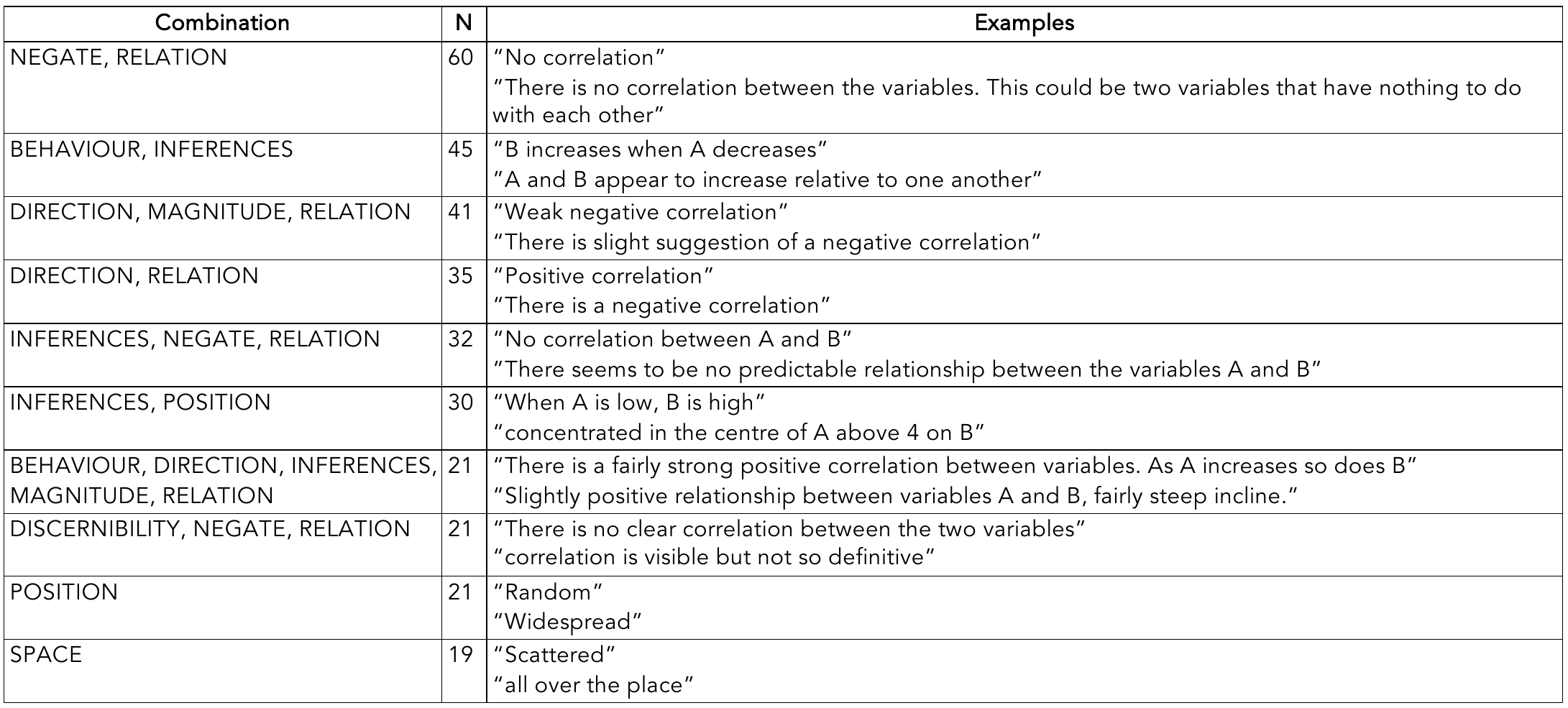}
\label{fig:top10comb}
\end{table*}

\noindent \textbf{Summarization and collocation analysis:} we began the exploration of results by looking at the global distribution of words, with the aim of finding out which words and parts-of-speech were prevalent in the data. The summary shown in \figurename~\ref{fig:topwords} reveals that a large number of answers made direct references to the variable names, followed by \textit{variable}, \textit{relationship} (a word that was part of the instruction), \textit{correlation}, \textit{increase}, \textit{high} and \textit{low}. In the figure, occurrences of words with different parts-of-speech appear together -- \textit{increase} includes both noun (singular and plural) and verb (third person singular or plural) forms. Nonetheless, we identified a high number of unique nouns compared to other parts-of-speech, which guided us to the next steps.\\


\noindent \textbf{Collocation analysis:} the high occurrence of nouns led us to investigate \textit{collocations}, which are word combinations that occur frequently throughout the data~\cite{manning1999foundations}. Here, it is important to note that our analysis had a different final objective compared to information retrieval scenarios where collocations are commonly used. In information retrieval, \textit{distinct} collocations are important to differentiate documents, while in our case the objective was not to characterize individual answers, but to identify aggregate themes. Initially, we turned our attention to collocations of nouns and adjectives in the form of \textit{bigrams} and \textit{trigrams}, which are collocations of two and three words, respectively. From the resulting list of bigrams and trigrams, we extracted a list of the most common adjectives and looked at their distribution across correlations (for more information, see the related notebooks in the online repository).


The analysis of the types of adjectives, and how often they appeared, provided an overview of the content of verbalizations across scatterplots. Although not many adjectives seemed unexpected, such as \textit{positive} and \textit{negative}, the variety of adjectives likely reflect the fact that the task procedure did not push participants towards particular characteristics of the relationship between the variables. Another observation in our analysis was the low occurrence of some potentially common collocations, such as "positive relationship" with only 22 occurrences, which suggested that collocation analysis alone was not enough to map the properties of visualization and data back to the responses. This first stage helped us achieve the first goal (\textbf{G1}), with an overview of the language constructs and the vocabulary used to describe the visualizations.



\subsubsection{Derivation of categories} 

In this step, our analysis methodology progresses from the automated, NLP-driven stages to a qualitative analysis process driven by human expertise and interpretation, with the collocation analysis serving as a starting point. To further organize the vocabulary, we separated adjectives and nouns into two lists of groups. In one list, each adjective was paired with a list of nouns that appeared next to it within collocations (e.g. ``strong'' paired with ``relationship, correlation, etc.''). The other list contains the inverse, with each noun being paired with a list of adjectives that appeared next to it in collocations. The same procedure was done also for adverbs and verbs.


This served as a starting point for developing the taxonomy. Due to the structured and limited nature of the textual data (i.e., narrow responses to experiment conditions rather than open-ended discussions), we followed a thematic analysis~\cite{braun2006using} methodology only partially. We treated the extracted parts-of-speech as \textit{codes} (as in thematic analysis) and developed a set of descriptive themes. As a first step, we make use of the existing distinctions between the parts-of-speech and organized the nouns and verbs under the \textbf{concept} category and the adjectives and adverbs, that modify the nouns and verbs, under the \textbf{trait} category. These became the two axes in the resulting taxonomy and directly reflect the language in which concepts are \textit{characterized} by the traits. In order to identify the sub-category names (i.e., themes) for concepts and traits, we base our attention on terminology commonly used in relation to data visualization and statistical relationships. This basis includes words that describe elements of charts and visual channels -- shapes, position, transparency, etc. -- and words related to data/statistics -- relationships, arrangements, characteristics of data.

Our objective was to define categories that are sufficiently distinct, but those that could also potentially encompass sub-categories and be open for future changes. The decision on the level of distinction was based on a subjective analysis of the words in potential categories or concepts: if the words that were grouped together were, overall, too different, then that concept was considered as too broad in scope. To arrive at the set of concepts and traits presented in this paper, one of the authors did an initial categorization of nouns, verbs, adjectives and adverbs, with the author's choice of categories. The second author then independently assigned the same categories to their own four lists. Rather than measuring level of agreement, open discussion and re-classification were used to refine the choices of \textit{concepts} and \textit{traits}. During this process, words that did not fit at all into the refined taxonomy were left without a category and deemed unhelpful for both the taxonomy and use in NLG applications, e.g., ``wild''.


This process led to the following \textbf{concepts}: \textit{relation}, \textit{behaviour}, \textit{graphic}, \textit{space} and \textit{inferences}. Table~\ref{fig:table-concepts} contains details about their definition with examples of words and representative responses to provide context. These concepts represent properties of the data (\textit{relation} and \textit{behaviour}) and visualizations (\textit{graphic} and \textit{
space}), as well as interpretative words (\textit{inferences}). As for \textbf{traits}, defined with examples in Table~\ref{fig:table-traits}, we arrived at the following: \textit{magnitude}, \textit{direction}, \textit{position}, \textit{discernibility}, \textit{regularity}. The traits are defined largely after adjectives and adverbs and represent how participants characterize the properties of visualizations and data, in terms of quantities and qualities.

\begin{figure}[!t]
\centering
\includegraphics[width=.7\columnwidth]{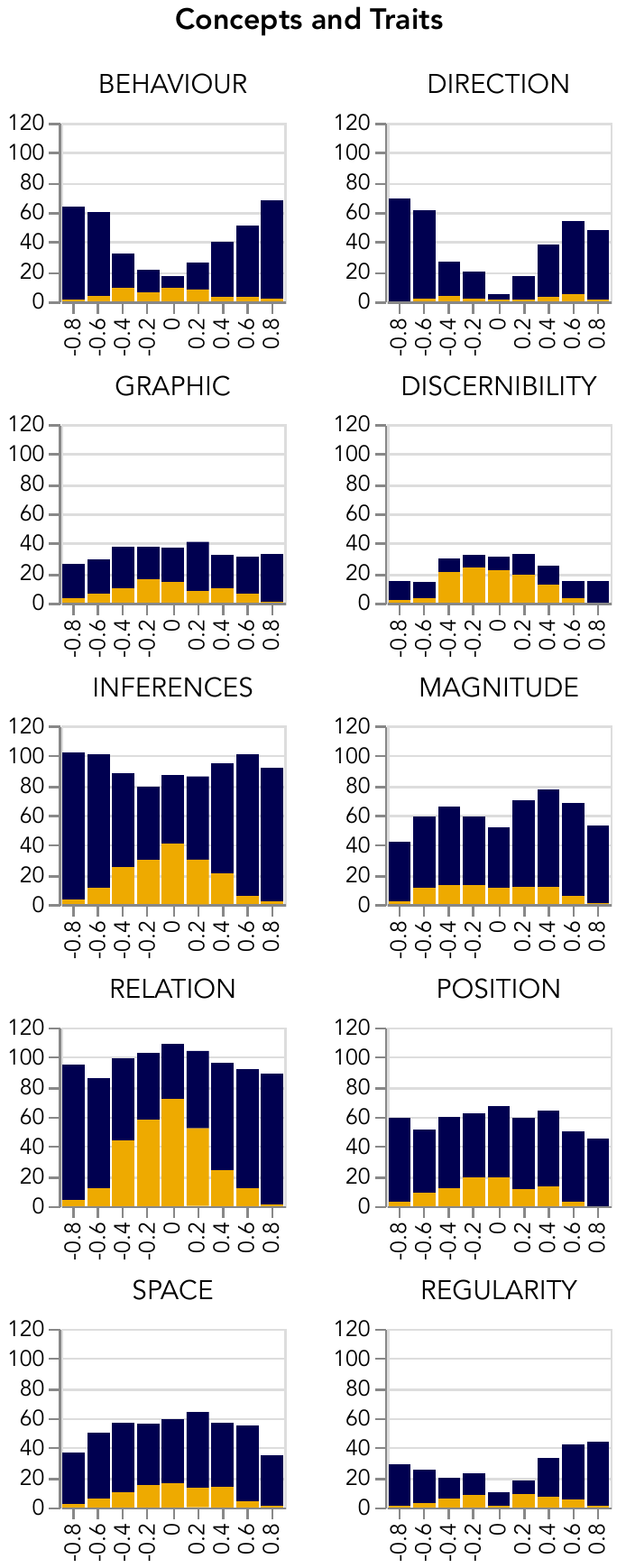}
\caption{Distribution of responses tagged with concepts and traits across scatterplots. The blue stacks indicate affirmative statements tagged with that concept or trait, while the gold stacks indicate the responses that were additionally tagged as a negative statement.}
\label{fig:dist-ct}
\end{figure}

We additionally defined a special category, \textit{negate}, to categorize responses between affirmative and negative statements, as a quick glance over the collected answers revealed the need to account for the negation of some of the concepts or traits, such as in ``no clear relationship''. This is visible in the distributions of the answers across scatterplots in \figurename~\ref{fig:dist-ct}.

\subsubsection{Response categorization and further refinement}

The process above resulted in four distinct lists of pairs of word and category. Lists of nouns and verbs were mapped to concepts, while lists of adjectives and adverbs were mapped to traits. Each response was tagged with a concept, trait or negation, based on looking up each word of the response in the lists. A response could have up to 11 tags, with tags not being repeatedly assigned to the same response (i.e., if a response had both ``weak'' and ``strong'', it was tagged with ``magnitude'' only once).  The tagging process also led to a further refinement of the categorized words: once a first round of tagging was done, we manually analyzed remaining uncategorized responses and added and categorized words. This was done until every response was tagged with at least one concept or trait. At this stage, a final set of 8 responses, out of the original 1428, were not tagged, as none of the words matched a concept or trait, e.g., ``Grocery shopping while hungry''. The refinement process considered the four main parts-of-speech -- nouns, verbs, adjectives and adverbs -- with the addition of prepositions that were not captured by collocation analysis, such as \textit{above} or \textit{below}. The complete set of tagged responses range from one concept to long, multi-sentence answers that include up to 9 concepts or traits. The completion of the taxonomy helped to partially complete our second goal (\textbf{G2}): identifying the characteristics of the data and the visualizations that participants refer to.




\begin{figure*}
\centering
\includegraphics[width=.9\textwidth]{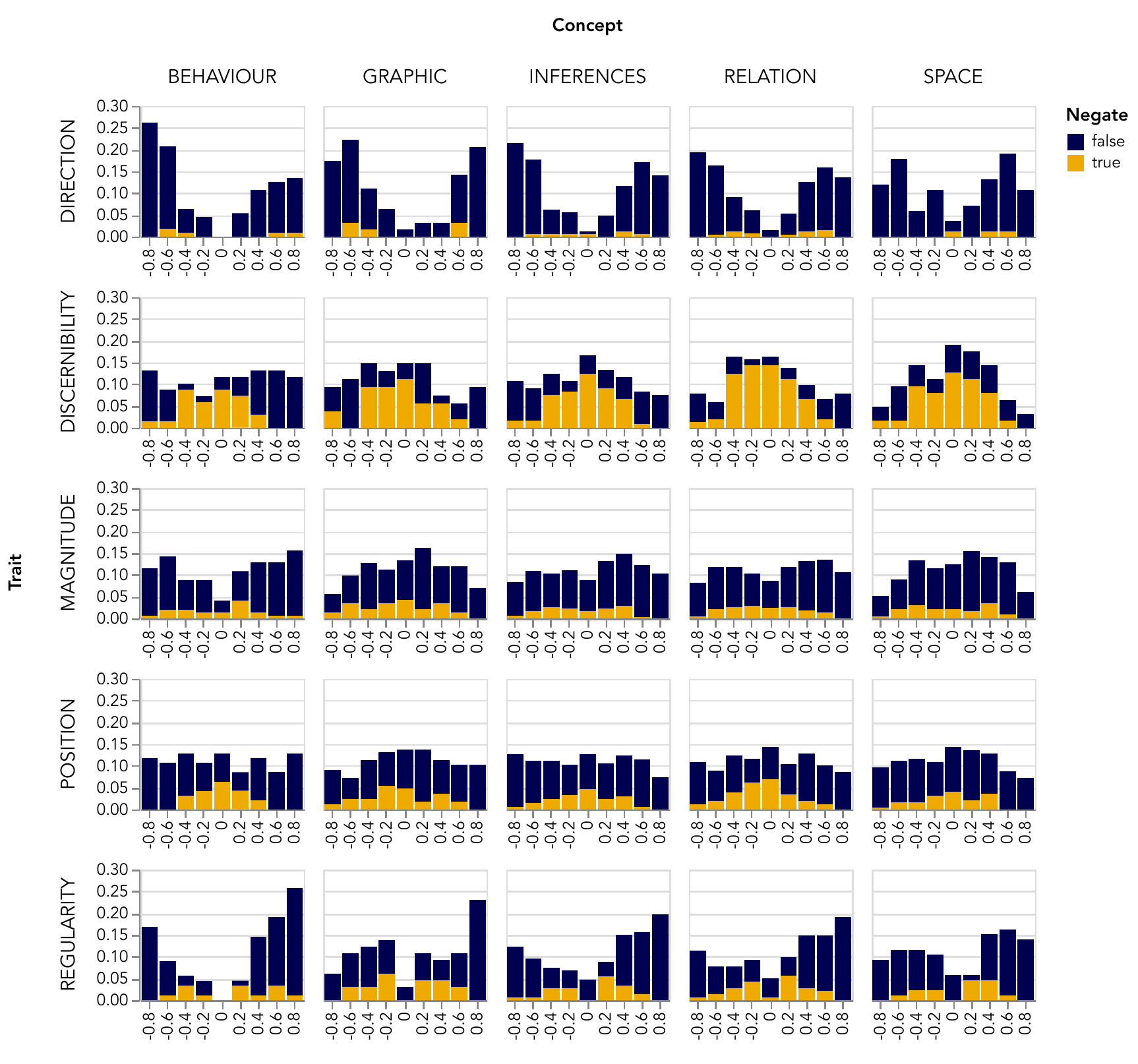}
\caption{Normalized distributions of the paired co-occurrence of concepts and traits in responses, per level of correlation, with color separating affirmative statements (blue) from negative statements (in gold). The chart shows the consistent patterns for the traits, and the dependency of the shapes of the distributions of concepts on the former. The traits that have a more uniform shape suggest that they are useful to describe scatterplots independent of the level of correlation, such as ``position'', ``discernibility'' and, to a certain degree, magnitude.}
\label{fig:cbyt}
\end{figure*}

\subsection{Findings from S1}
\label{sec:findings-s1}

The analysis of the tagged responses, with the taxonomy we derived, helps us reach the third goal (\textbf{G3}): understanding the relationship between the characteristics and the language observed in the studies. For an initial overview, we looked at the combinations of concepts, traits and negative statements for the whole data, revealing 419 distinct sets of combinations, with almost half of them (210) being unique to a single response. Table~\ref{fig:top10comb} shows the ten most common combinations of categories for the whole data; describing the \textit{absence} of relationship with a negation was the most common combination (e.g., \textit{``no clear relationship}''), followed by description of behaviour and inferences (e.g., \textit{``as A increases, B decreases''}). The wide diversity of descriptions of correlation in scatterplots is represented through the low number of responses with the most common combination (52), relative to the total number of answers (1428).


\figurename~\ref{fig:dist-ct} shows the distribution of concepts and traits across scatterplots, with the height of the bars depicting the absolute number of answers tagged with that concept or trait, and color distinguishing between affirmative and negative statements. For the concepts, the charts indicate a preference for describing the relationship between the variables (\textit{relation}) and the inclusion of the names of the variables (\textit{inferences}). For the traits, the variation of shapes for all distributions suggests that the level of correlation has an influence on how participants use those traits. As each response was assigned more than one tag, they might be counted in more than one chart in the figure.

\figurename~\ref{fig:cbyt} shows the normalized co-occurrences of the same concepts and traits across scatterplots, again with each stacked bar corresponding to affirmative (in blue) and negative statements (in gold). The same note from the previous chart applies here regarding the multiple counts of the same response across the matrix. The figure shows that for most combinations, there seems to be a variation driven by the level of correlation.  Two pronounced examples are \textit{direction} and \textit{behaviour}, which are more common in extreme values of correlation, i.e., a bi-modal distribution with peaks around higher and lower correlation values. This indicates that participants identify strong correlations with their inclinations, e.g., \textit{positive}, \textit{negative}, \textit{increasing} and how the points are arranged, e.g., \textit{trend}, \textit{pattern}, \textit{rate}. Others, such as \textit{magnitude} or \textit{inferences}, are used uniformly regardless of the correlation.

Such ``uniform'' categories indicate language that is regularly used by participants to describe the scatterplots. For instance, the magnitude of the correlations is often important to mention and participants indicate inferences and try to interpret the ``meaning'' of the scatterplots irrespective of the strength of relations. 
An increasing use of negative statements for position, combined with inferences and relations, for the lower levels of correlation, is also noticeable. An example utterance with this combination is \textit{``the relationship between A and B is random and there is no correlation''}. Discernibility also contrasts with the other traits due to the higher proportion of negative statements instead of affirmative (e.g., \textit{``there is no clear relationship''} being more common than \textit{``there is a clear relationship''}).

In addition to these prominent observations, we also identified what we consider peculiarities on how participants used some of the concepts. An interesting use of the concept of \textit{space} was in describing a coordinate system, with the variables A and B as references. Examples of such utterances are ``loose cluster of data points distributed around A6, B6 with a few wild outliers'' and ``scattered points concentrated most densely between 6A and 4-8B''. What surprised us is that the scatterplots do not contain gridlines, and ticks are included only for even numbers (which at least explains why most of the grid references are indeed for the even positions).





 \section{Study S2: Verbalization to Visualization}
\label{sec:e2}

In the first study, we asked participants to provide descriptions of visualizations, asking them to view scatterplots and \textit{verbalize} what they saw. In the second study, we aim to characterize the reverse: given a particular verbalization, participants were asked to choose the scatterplots that represented that verbalization. The results of this study indicate how much verbalizations resonate with participants for the various levels of correlation, represented through scatterplots. This study advances us towards the third goal (\textbf{G3}): understanding the relationship between these characteristics and the language observed in the studies. 

\subsection{Preparation}
\subsubsection{Materials}
\label{sec:materialprepE2}

For this study, we wrote 15 statements based on 7 combinations of concepts and traits. The combinations were chosen based on the coverage of concepts and traits from our taxonomy which enabled us to systematically vary the statements while maintaining sentence structure across combinations (\textit{"There is [trait] [concept] between A and B"}). We also aligned the statements to utterances commonly observed in the first study in order to ensure that the resulting statements would be widely understood by the participants. 

Four of the seven combinations include the \textit{relation} concept, as it is the most common, with terms such as \textit{correlation} and \textit{relationship}, combined with four traits: \textit{magnitude, direction, regularity} and \textit{discernibility} -- the last trait was additionally combined with a \textit{negation}. Another pair of statements was the combination of \textit{behaviour} and \textit{direction}, whilst the last two pairs of statements combine \textit{inferences} and \textit{space} with \textit{position}. This resulted in the following statements:

\textbf{(A) Relation and Magnitude:}

\noindent \textbullet\ \textit{There is strong correlation between the variables A and B}

\noindent \textbullet\ \textit{There is moderate correlation between the variables A and B}

\noindent \textbullet\ \textit{There is weak correlation between the variables A and B}

\textbf{(B) Relation and Discernibility (with Negation):}

\noindent \textbullet\ \textit{There is a clear relationship between A and B}

\noindent \textbullet\ \textit{There is no obvious relationship between A and B}

\textbf{(C) Relation and Regularity:}

\noindent \textbullet\ \textit{There is a tight relationship between A and B}

\noindent \textbullet\ \textit{There is a loose relationship between A and B}

\textbf{(D) Relation and Direction:}

\noindent \textbullet\ \textit{There is a positive correlation between the variables A and B}

\noindent \textbullet\ \textit{There is a negative correlation between the variables A and B}

\textbf{(E) Behaviour and Direction:}

\noindent \textbullet\ \textit{There is an upward trend}

\noindent \textbullet\ \textit{There is a downward trend}

\textbf{(F) Inferences and Position}

\noindent \textbullet\ \textit{When variable A is high, so is variable B}

\noindent \textbullet\ \textit{When variable A is high, variable B is low}

\textbf{(G) Inferences, Space and Position:}

\noindent \textbullet\ \textit{When the values of A are high, so are the values of B}

\noindent \textbullet\ \textit{When the values of A are high, the values of B are low}

\begin{figure*}[!tb]
\centering
\includegraphics[width=.85\textwidth]{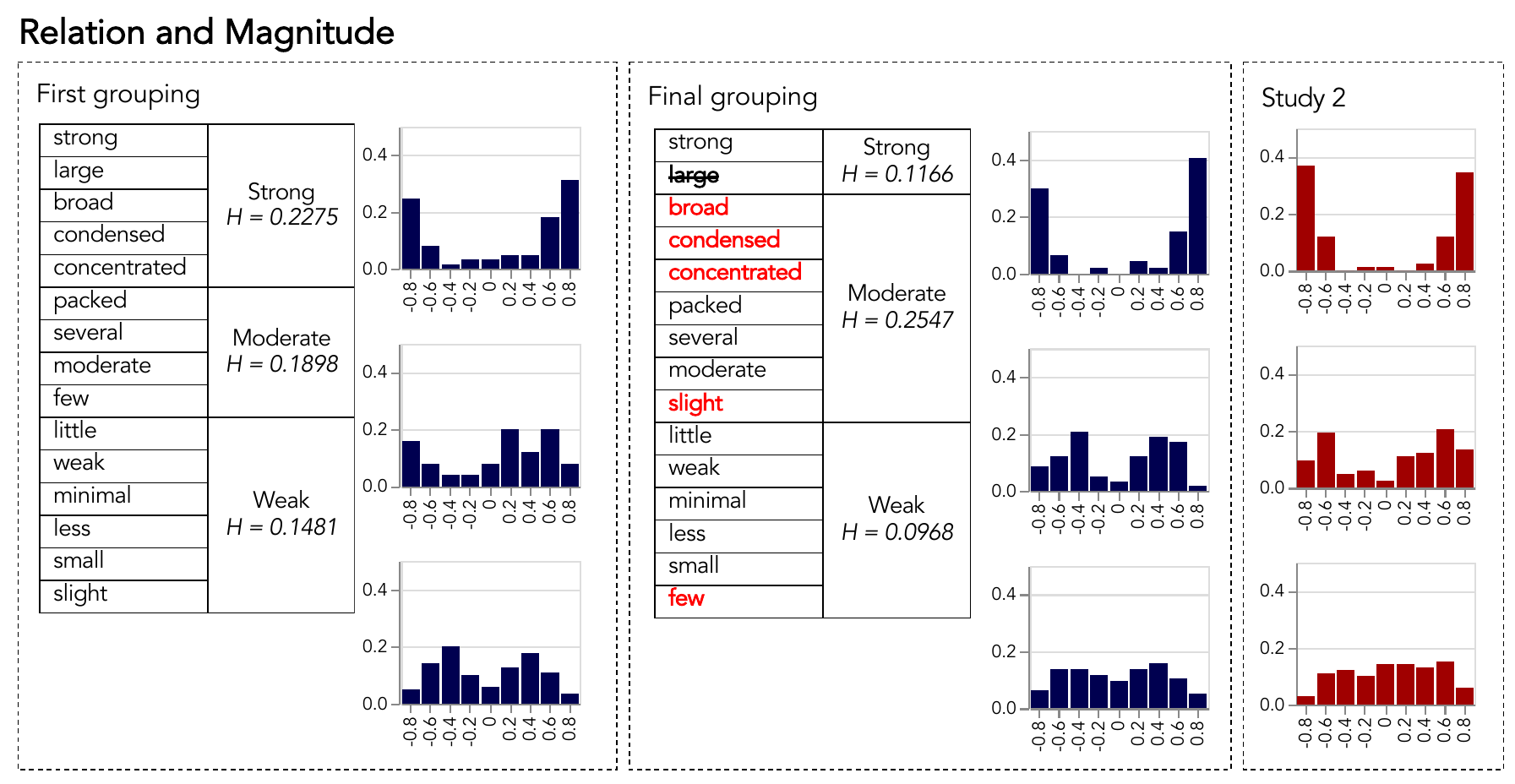}
\caption{Study 2 results for the Relation and Magnitude category (right, red bars) and the iterative steps of the \textit{semantic grouping} for the S1/S2 comparison. The first groups are defined based on the intensity-driven word vector distances, with acknowledgment that it is not the optimal configuration. After step one, words are moved between groups or dropped until the distances between the two studies, for each group, are subjectively acceptable. In the example, the best configuration of answers from S1 which are mapped to \textit{Strong} is when the word \textit{strong} itself is the only word left in the group, i.e., the configuration where the Hellinger distance between the distributions of S1 and S2, for \textit{Strong}, is minimum with the extracted vocabulary.}
\label{fig:magrel-steps}
\end{figure*}

\subsubsection{Participants and procedure}

We recruited 120 participants agian through the Prolific platform. Our pre-screening filter was self-reported English as a native language and not having participated in the first study. For the survey itself, we used the Qualtrics system\footnote{https://www.qualtrics.com/}. The scatterplots from S1 were positioned on a 3x3 grid and labeled sequentially \textit{Chart 1, \ldots, Chart 9} (see \figurename~\ref{fig:os2}). A randomizer was set up so that each participant would see 5 statements and thus each sentence would have been seen by a similar number of participants. Here we note that a first group of participants were given only 4 statements, however in the second, larger run, we decided to increase the number of statements shown to 5 after noticing that participants did not display any signs of fatigue. We kept all the answers from these runs as there were no discernible changes in responses.

Before proceeding with the task, participants were presented a training description on the use of scatterplots in correlation analysis in a similar setup as the first study (see the supplemental repository for details). In the task itself, each participant saw the instruction \textit{``Please select all the charts that apply to the following statement''}, followed by the corresponding statement and the grid of scatterplots (\figurename~\ref{fig:os2}). 

In this procedure, there are three points of discussion that could be potential concerns: participants' lack of knowledge about statistics, lack of attention to the task at hand or lack of engagement with the study. To address the first issue, we included one knowledge check question on correlation. We evaluated the response to this "test question" along with participants' other responses and did not filter any participants since all of them demonstrated a reasonable understanding of correlation. This decision was also reinforced by the fact that we are not judging accurate estimations and that some of the statements involved highly subjective interpretations. The answers to this knowledge check question are included in the complete data in the online repository. Regarding the other two concerns, we did not include further specific attention checks and did not notice major issues in the results. 

\subsection{Findings from S2}
\label{sec:s2resultsanalysis}
Overall, we collected 550 responses from the 120 participants. We first discuss and analyze the results for each statement. 
Here, we analyze the results with the semantic distinctions between the concept/trait combinations in mind. The first three combinations (A, B, C) involve terms for which the meanings are not directly linked to the underlying mathematical definitions or to the visual forms of the charts. The following four combinations (D, E, F, G) include statements that have a closer association with the mathematical definitions of correlation, such as direction with \textit{positive} and \textit{negative}. This means that for such statements, it is possible to assign a \textit{correct} answer. However, as our study was not designed to obtain accurate estimates, our analysis focuses on the interpretation of the statements in relation to the scatterplots and we refrain from making wide assessments on correctness or accuracy.

\begin{figure*}
\centering
\subfloat[ \label{fig:reldis}]{
\includegraphics[width=.33\textwidth]{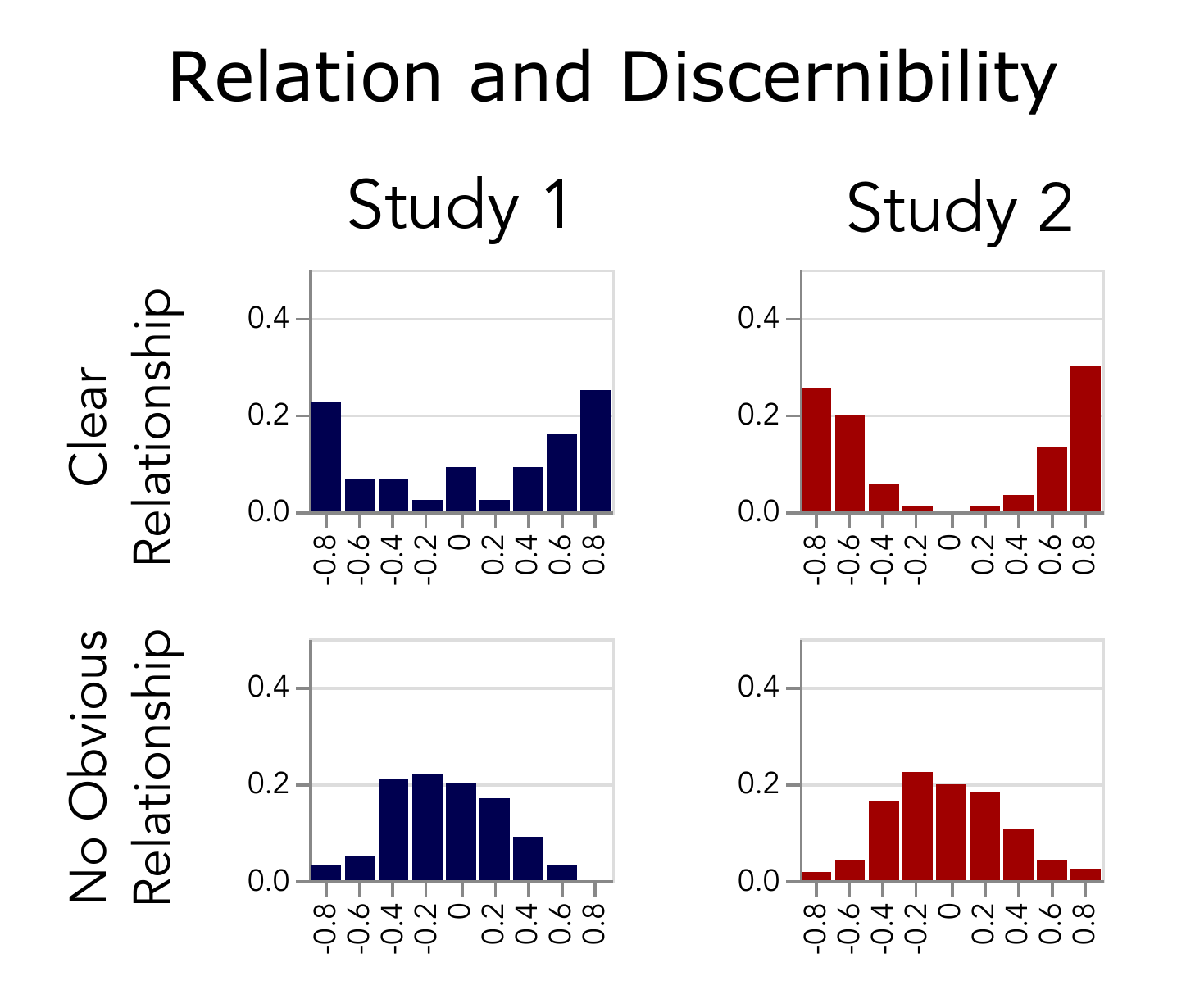}}
~
\subfloat[ \label{fig:relreg}]{
\includegraphics[width=.33\textwidth]{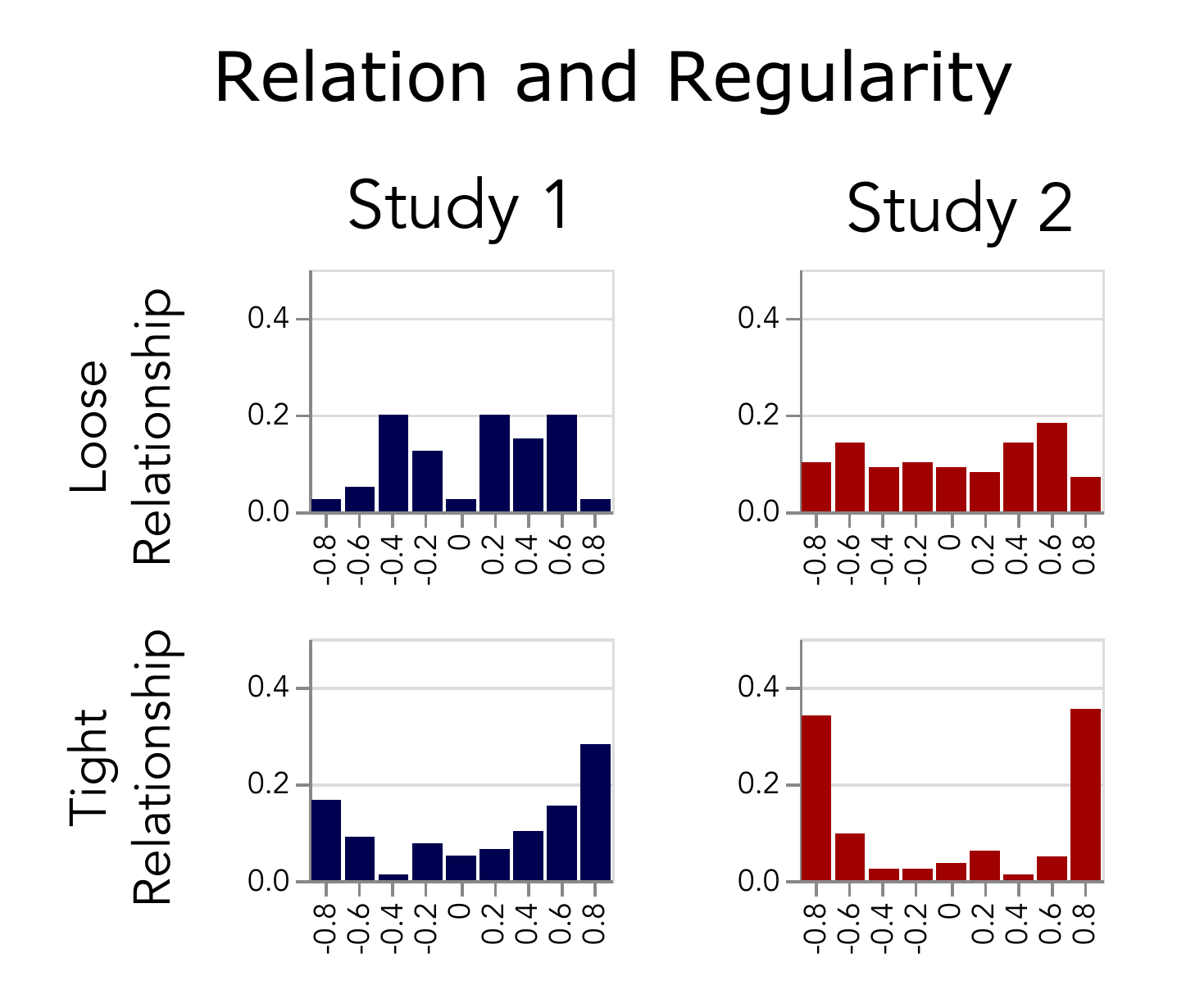}}
~
\subfloat[ \label{fig:reldir}]{
\includegraphics[width=.33\textwidth]{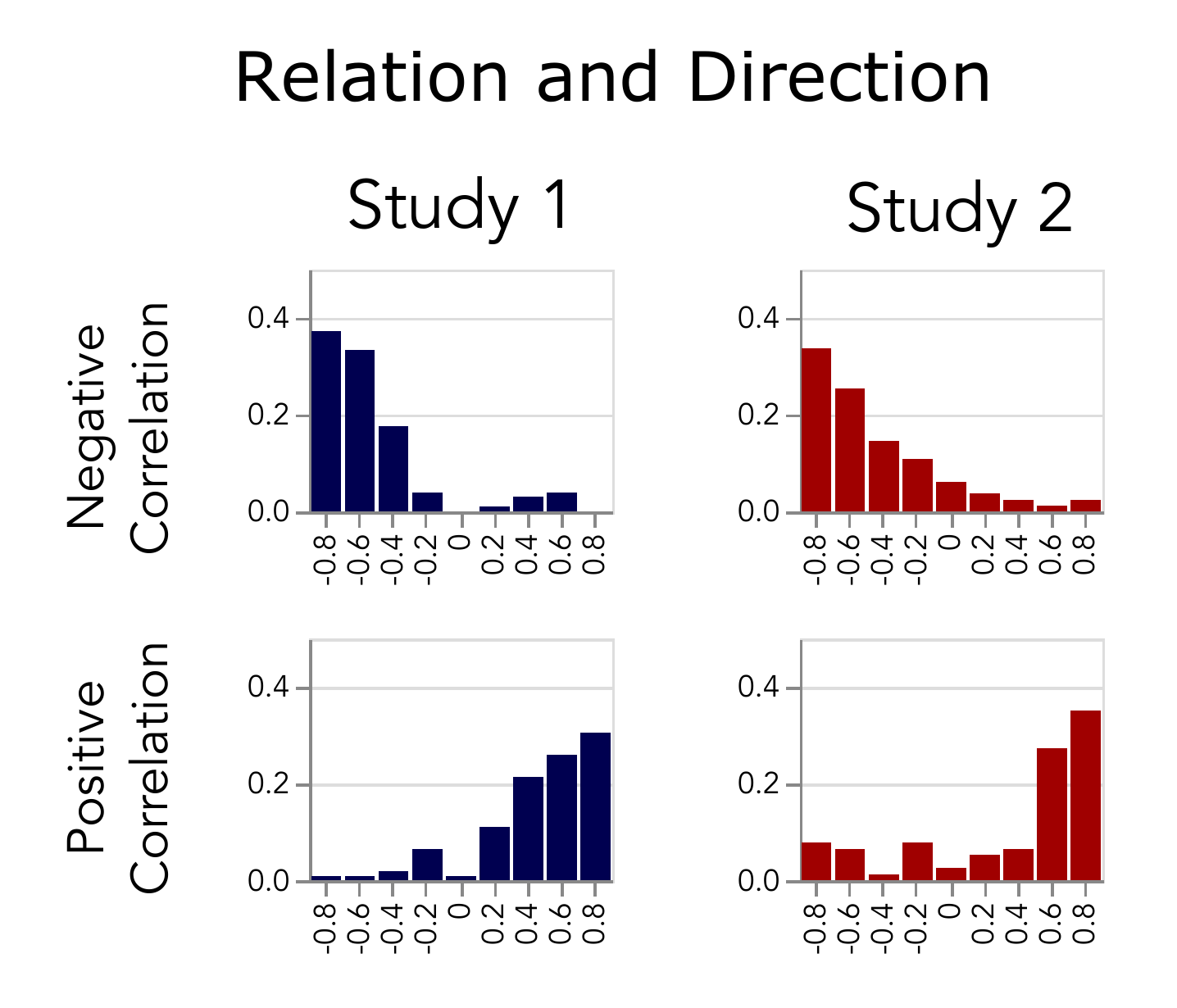}}
\\
\subfloat[ \label{fig:behdir}]{
\includegraphics[width=.33\textwidth]{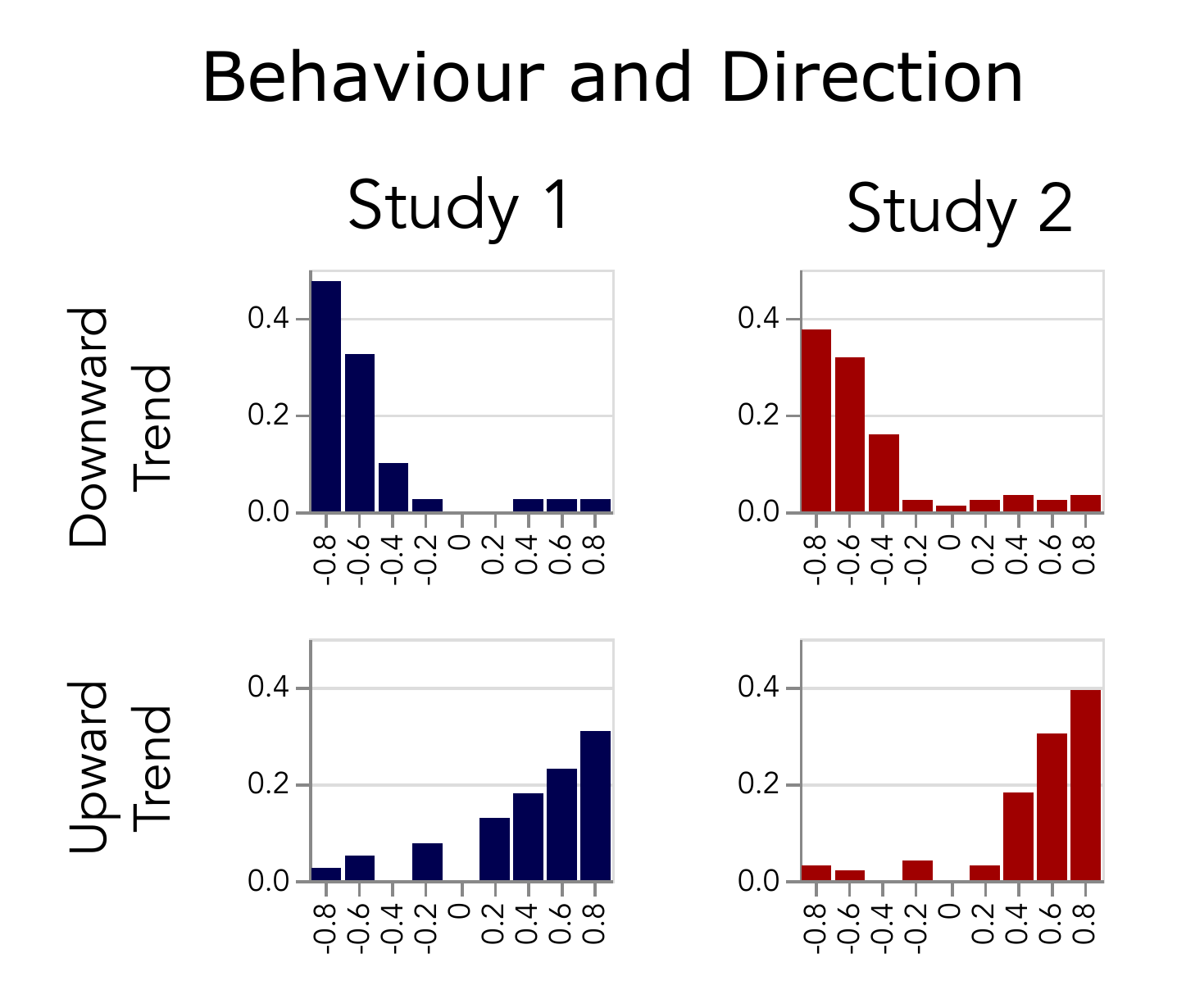}}
~
\subfloat[ \label{fig:infpos}]{
\includegraphics[width=.23\textwidth]{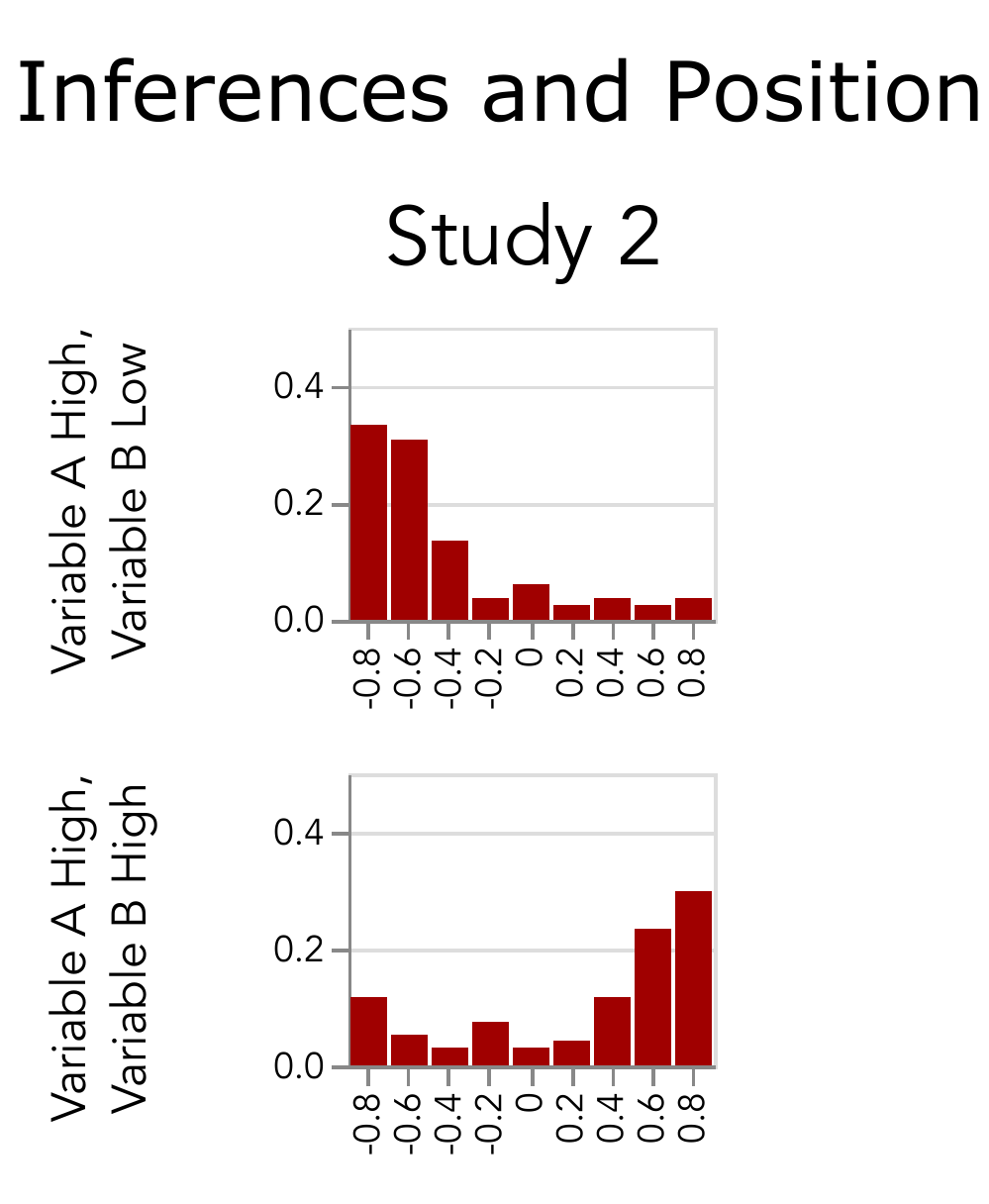}
}
\subfloat[ \label{fig:spacepos}]{
\includegraphics[width=.23\textwidth]{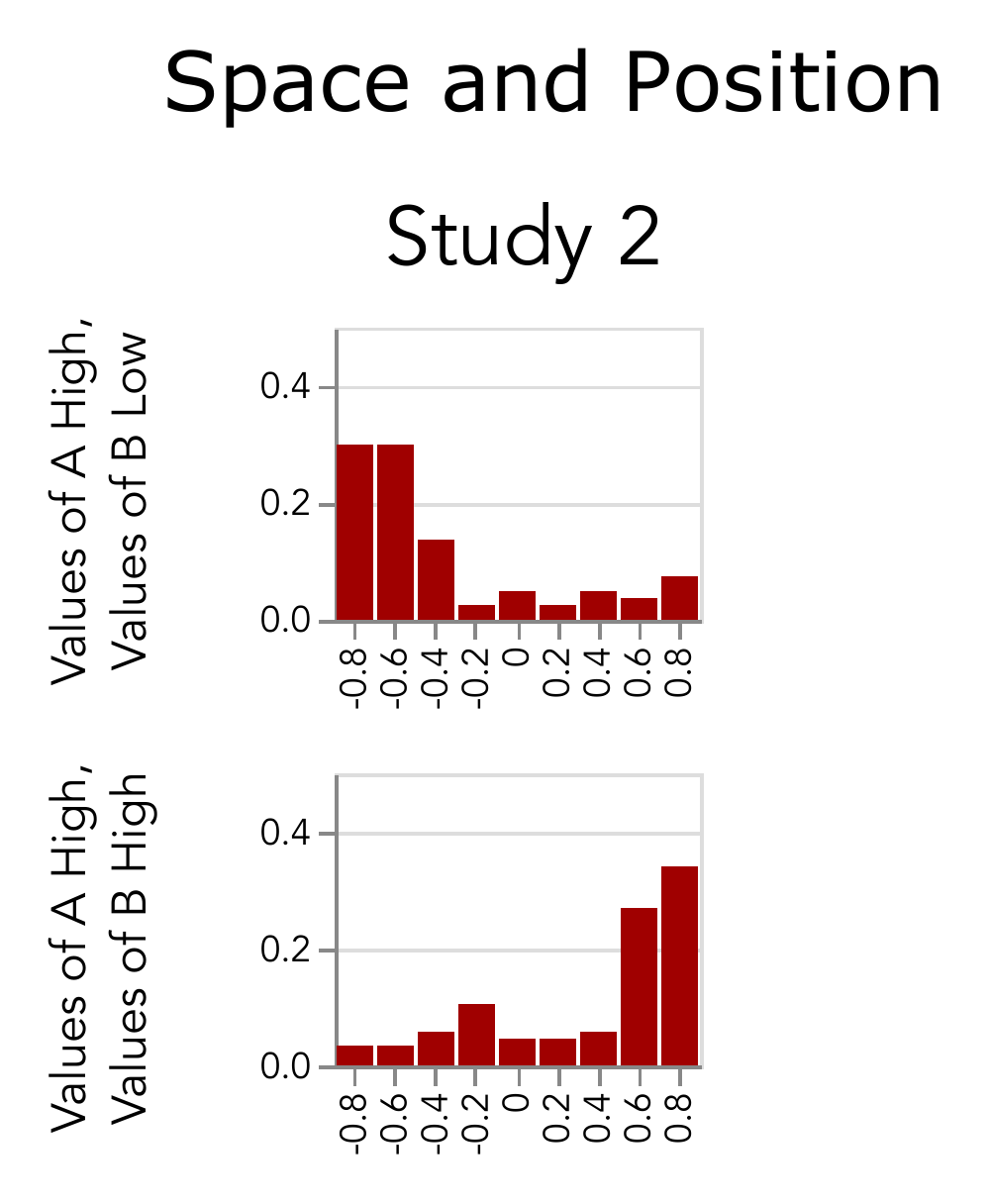}
}
\caption{Juxtaposition of results of Study 2, described in detail in~\Cref{sec:e2}, and the equivalent distributions of responses across scatterplots from Study 1, which are used in the comparison between the studies, described in~\Cref{sec:alignment}. The results of S2 for the last two combinations are not compared with S1 due to the different sentence construction, which makes it difficult for a simple comparison to be done in the same way as the other statements.}
\end{figure*}

\textbf{(A) Relation and Magnitude:} this category included three statements, with magnitude varying between \textit{strong}, \textit{moderate} and \textit{weak}. For \textit{strong}, the distribution of responses is bimodal and symmetric, with peaks in the extreme values (rightmost box in \figurename~\ref{fig:magrel-steps}), indicating that participants have an almost unambiguous association of \textit{strong} with minimum and maximum values of correlation. For \textit{moderate}, although the distribution is also bimodal and symmetric, it is not skewed and the peaks are located towards the mid-values of correlation. The shape of the distribution suggests that, when all scatterplots are seen together, \textit{moderate correlation} is more ambiguous. Lastly, the distribution for \textit{weak} is relatively uniform, except for the low occurrence in extreme values, indicating that even the absence of correlation is associated with the word \textit{weak}.

\textbf{(B) Relation and Discernibility:} this combination includes one \textit{affirmative} statement and one \textit{negative} statement. The results (red chart in \figurename~\ref{fig:reldis}) suggest that participants less ambiguously associate the existence of a \textit{clear} relationship with the scatterplots that depict higher levels of correlation (0.6 and 0.8, positive and negative) and, conversely, \textit{no obvious} relationship from -0.4 to 0.4.

\textbf{(C) Relation and Regularity:} the statements for this combination refers to \textit{tight} and \textit{loose} relationships. The distribution of results (red chart in \figurename~\ref{fig:relreg}) shows that participants strongly associated \textit{tight relationship} with the extreme values of correlation with little ambiguity. In contrast, the distribution for the choices associated with \textit{loose} is almost uniform, with very small peaks around $\pm$0.6, suggesting that it is an ambiguous term. Although we did not ask participants regarding their views on the meaning of words, we also speculate that \textit{tight} and \textit{loose} are not interpreted as having exact opposite meanings in the context of a data relationship such as correlation.

\textbf{(D) Relation and Direction:} this combination had two statements, for \textit{positive correlation} and \textit{negative correlation}. The distribution of answers (red chart in \figurename~\ref{fig:reldir}) for \textit{negative} is right-skewed, decreasing across correlations. For \textit{positive}, the distribution is left-skewed, being uniformly low from -0.8 to 0.4 and then sharply increasing at 0.6. We speculate that the differences between the results are likely the result of an anchoring or baseline effect, as participants saw all stimuli at the same time. However, further experiments would be needed to confirm such effect or clear up the differences between positive and negative. 


\textbf{(E) Behaviour and Direction:} for this combination, the statements mentioned either an \textit{upward trend} or a \textit{downward trend}. Unlike the previous combination, the results for both statements (red chart in \figurename~\ref{fig:behdir}) seem to be more closely aligned, being right-skewed for downward and left-skewed for upward in a similar manner. We speculate that the difference between the categories is due to the fact that \textit{upward} and \textit{downward} are associated with the visual form rather than the mathematical definition of correlation.

\textbf{(F) Inferences and Position:} for this combination, the focus is on simply describing position (\textit{high} and \textit{low}) without any additional traits or concepts (besides the references to the variables). There is therefore a stronger tendency on interpreting the visuals rather than the correlation relationship. As seen in \figurename~\ref{fig:infpos}, participants selected scatterplots mostly in the range of $\pm$0.6-0.8. 
For this category, all scatterplots were selected at least once, which is unexpected since some of the scatterplots do not match the statement at all. We speculate that the use of two clauses in the statement might have made some participants disengage with the task.

\textbf{(G) Inferences, Space and Position:} this combination is very similar to the previous one, but includes the word \textit{values} instead of \textit{variable} in the test statements. The intention was to provide a more concrete reference to the correlation relationship through the word \textit{values}. Results, however, are very similar to the previous combination, with the selections within the same range of $\pm$0.6-0.8 (see \figurename~\ref{fig:spacepos}). Once again every scatterplot was selected at least once and our speculation is the same regarding the disengagement or confusion. As we did not ask for justifications in the study or followed up with participants, this has not been fully clarified.

\section{Comparing S1 and S2}
\label{sec:alignment}
Due to the complementary nature of the two studies, we do a comparative analysis to see if how the interpretation of concepts and traits vary between \textit{visualization to verbalization} and \textit{verbalization to visualization} settings. This analysis strengthens our observations towards the third goal (\textbf{G3}): understanding the relationship between the characteristics and the language observed in the studies.

In order to compare the studies, we must first define what exactly is being compared based on the the kind of data that was collected. In S1, we can count -- per level of correlation -- the aggregate number of answers containing the words for each concept, trait, and their combinations. This forms a distribution of how each combination is represented across scatterplots and gives insight about the level of correlation at which terms might not resonate with participants anymore. This is also exactly what was obtained in S2, through a different study procedure. Thus, we can evaluate the alignment between studies and refine the vocabulary by comparing the distributions across scatterplots of the studies. However, aggregating the results from S1 still requires two preparation steps, due to the varied nature of concepts and traits.

\textit{Semantic grouping}: as mentioned previously, traits such as \textit{magnitude} contain words denoting different levels of intensity. In S2, the statements for the combination of \textit{relation} and \textit{magnitude} included the words ``weak'', ``moderate'' and ``strong''. However, the vocabulary acquired in S1 includes several words that need to be grouped with one of these three words. This is valid for both the comparison and for transferring the vocabulary into an NLG system. While ``weak'' and ``strong'' are easily distinguishable, it is challenging to define if ``slight'' matches ``moderate'' or ``weak''.

This step of grouping words can be done through a manual expert-driven or data-driven method. In this paper, we use a data-driven method as an example, based on an existing vector space of words~\cite{Kim2016a} where the distance between adjectives takes the semantic intensity of the adjectives into account. In this space, vectors representing words such as \textit{weak} and \textit{strong} have a longer distance between them than in unweighted vector spaces. The steps of this data-driven method are included in a Python notebook in the repository.

\textit{Affirmative-negative matching}: the other required step relates to binary traits, such as discernibility (``clear'' or ``vague''). In these traits, affirmative terms such as ``there is a vague relationship'' has a similar meaning to a negative term ``there is no noticeable relationship''. Since we used pairs of statements that include both affirmative and negative terms in S2, the distributions of answers from S1 must also be combined in a similar way. Further down this section we show examples of this step and the previous one, when comparing the results of the studies.

To quantitatively compare the results of two studies, a similarity metric is needed. From S1, we have answers aggregated into groups (in line with the statements of S2 as discussed above) along with frequencies of occurrence per level of correlation. From S2, we have a distribution of correlation levels chosen for each statement. As these distributions are discrete (quantities per level of correlation), we use Hellinger distance~\cite{Basseville2013} to calculate the similarity between matched normalized distributions from S1 and S2, for a specific combination of concept and trait (e.g., \textit{relation and magnitude}). The Hellinger distances are between 0.0 and 1.0, where values closer to 0.0 mean that the distributions are more similar and values closer to 1.0 mean that the distributions are more different. The metric is nonparametric and distribution-free, therefore it does not impose any assumptions about the shape of the distributions that are being compared. We note that we use the Hellinger distances as heuristics to compare and comment on the similarities between S1 and S2 for the different concept/trait combinations, rather than a definite measure providing thresholds for choosing particular words in the vocabulary.

\subsection{Comparison for refinement}

\textbf{Relation and Magnitude:} as described above, the semantic grouping step is required to be able to compare the results of the \textit{magnitude} trait between studies. Here, the calculation of the Hellinger distance provides a guidance for changing the semantic groups iteratively. Figure~\ref{fig:magrel-steps} illustrates the initial and the final steps of the process, juxtaposed with the results of S2. ``First grouping'' shows the semantic grouping based on the aforementioned vector space (see the external repository for further details), with words from S1 mapped to the three words used in S2. The ``final grouping'' (intermediate steps not shown for the sake of brevity) shows the final results after a few iterations, with words moved between groups or removed altogether. In this example, this stops after the distances are not getting any smaller or there are no further candidates to be moved between groups.

The comparison of results indicate that \textit{strong} is a suitable category, aligned between studies, but it is also best represented by the word strong alone, without other potential synonyms found in S1. \textit{Moderate} also aligns well between studies and includes some interchangeable words. \textit{Weak} aligns well too, but is quite ambiguous for mid-range and lower levels of correlation.


\subsection{Comparison for alignment}




\textbf{Relation and Discernibility \& Relation and Regularity:} the comparison of these two pairs (see \figurename~\ref{fig:reldis}) aims at giving insight into how the traits vary between the two scenarios --  \textit{visualization to verbalization} and \textit{verbalization to visualization}. For \textit{discernibility}, from the answers of S1, statements were matched to \textit{clear} or \textit{vague}; for \textit{vague}, negated statements for \textit{clear} were also grouped together. The Hellinger distances between the distributions are 0.2917 for \textit{clear relationship} and 0.1260 for \textit{no obvious relationship}, validating the choice for the traits. The slightly higher distance for \textit{clear relationship} suggests that the \textit{absence} of a relationship is a stronger concept than the \textit{existence} of it.

For \textit{regularity} (see \figurename~\ref{fig:relreg}), the distributions of answers from the first study were combined in the following manner: responses mapped to ``loose'' were combined with the negated responses for \textit{tight}, whilst answers mapped to \textit{tight} were combined with negated responses for \textit{loose}. The Hellinger distances of 0.2554 and 0.2514 for \textit{loose} and \textit{tight}, respectively, indicate the good alignment between the studies. However, as discussed above, a \textit{tight} relationship is a much less ambiguous concept than a \textit{loose} relationship in the verbalization to visualization scenario. As the second study only covered the word ``loose'', further experiments focused on linguistic aspects would help to clarify the differences.

\medskip

\textbf{Relation and Direction \& Behaviour and Direction:} For these two categories, the comparison is again aimed at understanding the differences in the scenarios. For the first combination, the distances between the distributions are 0.2509 for \textit{negative correlation} and 0.2444 for \textit{positive correlation}. The main difference between the results (see \figurename~\ref{fig:reldir}) is the shape of the distribution for positive correlation -- S2 responses are concentrated on 0.6 and 0.8, whereas in S1, participants identified the scatterplots with as low as 0.2 correlation as positive. 

For behaviour and direction (see \figurename~\ref{fig:behdir}), the distances are slightly lower, with 0.1555 for \textit{downward trend} and 0.1691 for \textit{upward trend}. As discussed in the previous section, it is possible that the small difference between the two categories relates to the fact that upward and downward are linked to the visual form, but the data collected do not support further conclusions. From the results, both combinations seem appropriate for describing correlation in scatterplots.

We did not compare results for the last two statements -- \textit{inferences and position} and \textit{space and position} -- due to the different sentence structure, which does not provide a good baseline for aggregating results from S1.

\section{Discussion}
\label{sec:discussion}

\subsection{Reflecting on the results and implications}
The outputs of our studies are the taxonomy of concepts and traits, the distribution of categorized responses from S1 and the distribution of selected scatterplots from S2, as well as the observations reported from the analysis of the study data and their comparisons. The taxonomy and distribution of responses indicate a wide range of vocabulary that participants use to refer to and describe visual and data properties. Concepts and traits are used differently for the various correlation levels, with behavioral and directional aspects being more common in the extreme levels, for example. "Magnitude" category -- which usually expects participants to relate to the concept of correlation, e.g., "strong/weak correlation", used uniformly irrespective of the strength of correlation but used less for stronger correlations when compared to the "direction" category that usually relates to visual structures. This can potentially indicate that viewers are relating to the visual patterns more easily instead of thinking about the concept of correlation -- one  would, of course, need to run a targeted study to confidently claim the above.
We also observe that responses often involve multiple concepts and traits with particular combinations being more common. For instance, ``behaviour'' and ``direction'' (e.g., downward trend) are more common for higher levels of correlations and almost absent for lower levels.

The comparison between studies indicate many similarities between the processes of \textit{visualization to verbalization} and \textit{verbalization to visualization}; however, words from the extracted vocabulary are not always interpreted in the same manner for these processes. This is especially relevant for traits where adjectives and adverbs are not binary in nature. This is the case, for example, for \textit{magnitude}, which can indicate several levels of intensity with words such as \textit{weak}, \textit{moderate}, \textit{slight}, \textit{strong}, among others. The comparative analysis in Section~\ref{sec:alignment} enabled us to refine the vocabulary, narrowing it down to a set of synonyms that are well understood in the two settings studied. While tools that use natural language as input can ask for user input for corrections of meaning~\cite{Hoque2017}, NLG-based systems would benefit from knowledge that indicates whether \textit{slight} can be used interchangeably with \textit{weak} or \textit{moderate}. 

\subsubsection{Practical implications of this paper}

Here we discuss some immediate implications of our results for researchers and designers working on and with multimodal systems:

\textit{In designing multimodal systems:} The vocabulary captured and organized by the presented taxonomy could be considered a part of what Mitchell et al. refer to as ``dialogue seed corpus''~\cite{mitchell2014crowdsourcing}. Designers of multimodal systems could adopt this vocabulary for generating consistent pairings of scatterplots and their verbal descriptions. Our outputs also support the challenges outlined in the natural language generation (NLG) literature regarding the description of visible objects~\cite{Krahmer2012,Mitchell2013}, which in our case are data visualizations, as well as the challenge of deciding \textit{preferred} vocabulary in NLG tools~\cite{Hervas2013}.

\textit{In designing visualizations:} The differences in how certain traits or concepts are observed over correlation levels provide indicators to which characteristics of data are more relevant when visualizing such data. This knowledge, in turn, can inform how scatterplots could be re-designed to account for these variations. For instance, as discussed earlier, we observe that \textit{direction} of patterns is more prominent than the \textit{magnitude} of the relation for higher levels of correlation -- this information can motivate designers to augment the scatterplots with visual embellishments that emphasize the direction of the relation further.

\textit{In studying multimodal representations:} The study protocol and the analysis process here provide a methodological blueprint on how the alignment between verbal and visual representations could be studied systematically. Researchers could adopt the protocol and adapt the supplemental analysis code to semi-automatically analyze the study data.

\subsection{Further discussion and limitations}
\textbf{Study design:} we took various decisions when designing the studies that directly affect the taxonomy and collected answers. From the first study, two important decisions are the generation of stimuli with abstract variables in the plots and the choice of participants. The vocabulary we acquired was directly affected by which references could be used by participants, both based on their previous experiences with data analysis and visualization and the features in the scatterplot. We recognize that some of the observations and language would be different for cohorts with varied numeracy skills, similarly to how expertise affects how visual patterns are perceived~\cite{xiong2019curse}. In future work that expands the taxonomy for specific types of visualization or that targets a narrower set of participants, a more rigorous screening procedure would be welcome. Real variables could also could also influence participants include additional interpretations of the scatterplots. 

For the second study, an alternative design that would elicit a similar kind of information is matching statements to scatterplots via a proxy measurement, such as level of agreement. While this works well for correlation estimation (such as in the studies by Rensink~\cite{Rensink2011}) and can potentially provide more nuanced results, our preference was to ask for direct mappings between the statements and the scatterplots, since we were interested in a format of results that would enable the comparison between the two studies.
A related question is about the viability of a setup where participants are given a list of selected words and asked to enter a number for the correlation strength indicated. Unlike the notion of probability, where there is a more widely understood link between the concept and the number~\cite{kentsherman1964}, the concept of correlation is open to interpretation. To further complicate matters, there is even an inconsistency between the different numerical metrics that measure correlation~\cite{hauke2011comparison}. Therefore, we decided against asking participants to represent their interpretations through a number or a level of confidence.




\textbf{Scope of the concepts and traits:} The concepts and traits that we derived from the results are limited by the context of \textit{correlation in scatterplots}. Although it is likely that, as high level ideas, they will generalize to other visualizations and properties of data, it is uncertain whether the same words would be used in different experimental setups. Nonetheless, as previously discussed and further outlined in Section~\ref{sec:roadmap}, our aim is to provide a broad baseline for future studies, which could \textit{extend} or \textit{replace} parts of our overall categorization and word mappings.

\textbf{Employing a semi-automated approach:} 
Our analysis methodology in this paper involves natural language processing and machine learning techniques and also interpretative qualitative methods.
One could consider to solely adopt qualitative methods such as thematic analysis~\cite{braun2006using} or content analysis~\cite{Vaismoradi2013} where the whole set of responses are coded and categorized manually by experts. With crowd-sourced studies, however, conducting such  analysis on the whole set is much less feasible due to the large volume of responses. To that end, the automated steps reduce the volume of data to manageable levels for manual analysis. Also, the data we operate on in our setting is much more structured and responses are much more targeted, making it suitable for automated analysis such as part-of-speech tagging or collocation extraction. 
We argue that a semi-automated approach brings together the strengths of these both worlds -- the use of automated approaches enables the construction of a taxonomy at a larger scale relative to purely qualitative methods; on the other hand, the human interpretation brings the expertise required for extracting and organizing the meaning of words in the context of visualization and properties of the data.

\section{Research Roadmap}
\label{sec:roadmap}

The research presented in this paper is a starting point towards the wider goal of understanding how natural language and visualization work together in data analysis contexts. In this section, we present a research agenda to facilitate the further exploration of this topic, for which the study design, computational analysis routines, and the resulting data will serve as stepping stones. 

\subsection{Verbalization as an evaluation methodology}
The studies that we describe provide a variety of insights on how participants internalize the information conveyed by the scatterplots and how they then externalize this information through language. Systematically conducted verbalization studies have the potential to serve as a new lens to look into visualization designs directly, providing insights into how they work and how well they meet the intended design goals. Visualization researchers already work with natural language utterances of people using their designs, collected through methods such as think-aloud protocols or interviews~\cite{carpendale2008evaluating}. However, the scope of language used in data from such studies is broad and requires extensive analysis and interpretation from researchers. We argue that visual-verbalization studies will complement existing visualization evaluation methodologies, by providing researchers with systematically structured data on the language used by people reading and processing information that is communicated through visual representations of data. This should eventually point out to the limitations and the strengths of visualization designs. A future challenge to address here is to conduct such studies for more complex tasks while still maintaining the applicability of the natural language processing techniques as described in this paper. 

Another potentially interesting avenue is to consider interaction, and how researchers can gather utterances from users that inform them on how well the interactive process is running in a visualization solution. One direction could be to investigate multimodal interaction interfaces, such as chatbots, as an experiment medium to gather data for the evaluation of  interaction between systems and users.


\subsection{Moving towards interactive systems}
One motivation for this work is the increasing prominence of interactive, multimodal data analysis systems~\cite{Setlur2016, Srinivasan2017a}. In this paper, we produce a taxonomy and empirically evidenced descriptions to help structure such multimodal interfaces. Our approach here, however, is limited to one-off utterances, i.e., not part of a dialogue but rather as responses to a single stimulus. The structure of dialogues and the interactive narrative flow are also key for the design of multimodal interfaces~\cite{liu2010collective}. A future direction is to extend visualization and verbalization studies to interactive settings where the sequential nature of interactive dialogues could also be captured.


\subsection{Linguistic analysis}

Although our approach employs NLP methods, linguistic aspects related to sentence construction or the association with the varied backgrounds of participants were not examined. This perspective can be part of a deeper investigation about the cognitive processes related to verbalization in data analysis contexts. Such an in-depth understanding, if related to the theoretical frameworks emerging from cognitive science in similar vein to what Padilla et al. did within the context of decision-making~\cite{padilla2018decision}, has the potential to inform designers on respecting and acting on cognitive limitations and strengths of viewers of visualizations.

\subsection{Visualization and data property path}

Immediate follow-up experiments, based on varying the visualizations or the underlying data property, would greatly enrich the taxonomy. Systematic variations of visualizations based on visual variables (e.g. color) can also be used to compare results with the baseline \textit{correlation in scatterplots} taxonomy. For NLG-assisted visual analysis, acquiring knowledge about how concepts and traits overlap or differ across taxonomies will help to generate statements that can be generalized to different types of visualizations.

Follow-up experiments can also vary the data property under focus or characteristics of the data, such as those characteristics governing cluster separation~\cite{sedlmair2012taxonomy}. Proxy measures for scagnostics~\cite{Wilkinson2005a} can be used to simulate data for further studies. As correlation studies are generally focused on bivariate datasets, investigating multivariate or univariate data, as well as temporal or spatial attributes, would also greatly expand the scope of the taxonomy and the potential for informing a wider range of NLG-supported tools.

\section{Conclusion}
We described two studies aimed at developing an understanding of which characteristics of data and charts are relevant when people process data visualizations. We contribute with data, analysis and insights towards that goal. Through the \textit{visualization to verbalization} study, we developed a taxonomy and associated responses to the categories. Our findings present a diversity of vocabulary for the different levels of correlation, with some concepts more prevalent than others in different ranges. We also show that our organization of concepts into utterances also resonates with users in a \textit{verbalization to visualization} study. 


We argue for a central role for systematic natural language experiments as a means of evaluating visualizations and advancing multimodal systems. The potential success of future work relies heavily on a multi-disciplinary thinking with concerted effort from researchers in visualization and linguistics, as well as cognitive scientists bringing in theories of cognition. This strand of research is likely to open up new directions and advance the prevalence of human-data interaction systems in our increasingly data-rich society. 

\ifCLASSOPTIONcompsoc
  \section*{Acknowledgments}
\else
  \section*{Acknowledgment}
\fi

We thank Johannes Liem for helping to set up the online studies and discussions and Radu Jianu for helpful comments. This work was supported by the UK Engineering and Physical Sciences Research Council (EPSRC) with grant number EP/P025501/1.

\bibliographystyle{IEEEtran-noulr}

\bibliography{nlVis-WordsEstimative}

\begin{IEEEbiography}
[{\includegraphics[width=1in,height=1.25in,clip,keepaspectratio]{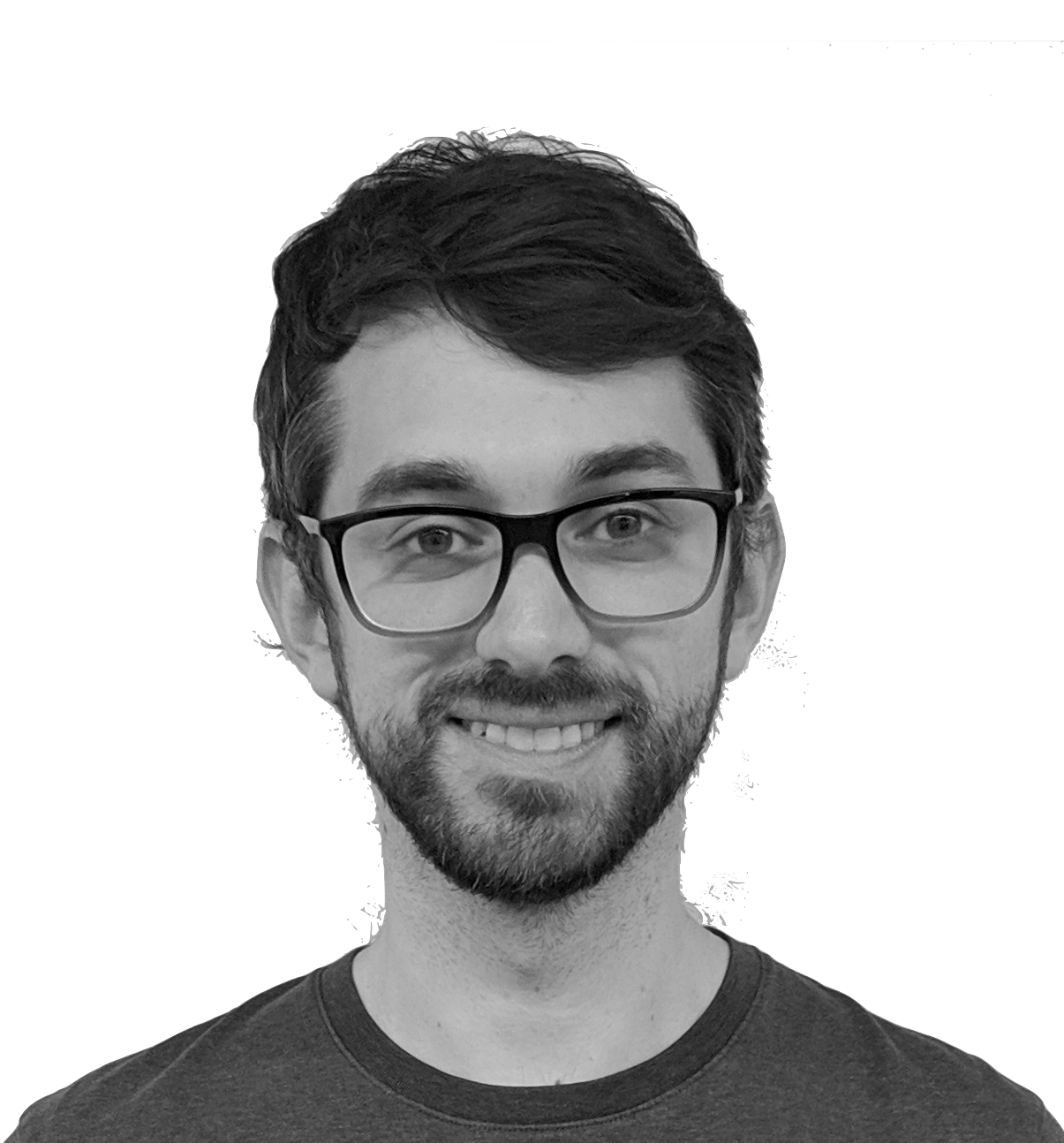}}]{Rafael Henkin}
    received a PhD in Computer Science from City, University of London. He is currently a Researcher at the Centre for Translational Bioinformatics at Queen Mary, University of London. The work presented here was done when he was a Research Associate in Visual Data Science at the giCentre at the Computer Science Department of City, University of London. He is interested in investigating methods that enable integrating interactive visualizations during data analysis, from a design perspective. He has been a reviewer for journals such as \textit{IEEE Transactions in Visualization and Computer Graphics} and conferences such as IEEE VIS and INTERACT.
\end{IEEEbiography}
\begin{IEEEbiography}
[{\includegraphics[width=1in,height=1.25in,clip,keepaspectratio]{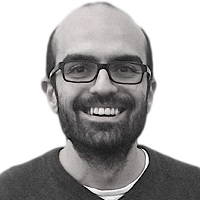}}]{Cagatay Turkay}
	is an Associate Professor at the Centre for Interdisciplinary Methodologies at University of Warwick. 
	His research focuses on designing visualisations, interactions and computational methods to enable an effective combination of human and machine capabilities to facilitate data-intensive problem solving. He serves as a committee member for several conferences including VIS and EuroVis, and part of the organising committee for IEEE VIS on 2017, 2018 and 2019. He served as a guest editor for \textit{ACM Transactions on Interactive and Intelligent Systems} and \textit{IEEE Computer Graphics and Applications}, and an editorial board member for \textit{Computers and Graphics} and \textit{Machine Learning and Knowledge Extraction} journals.
\end{IEEEbiography}

\vfill

\end{document}